\documentclass[useAMS,usenatbib,usegraphicx, usedcolumn]{mn2e}
\usepackage{color}
\usepackage{amsmath}
\usepackage{wasysym}
\usepackage{aas_macros}
\begin{document}
\bibliographystyle{mn2e}
\title[The $ugrizYJHK$ luminosity distributions]
{The $ugrizYJHK$ luminosity distributions and densities from the combined MGC, SDSS and UKIDSS LAS datasets.}

\author[Hill et al.]
{David T. Hill$^{1,2}$\thanks{E-mail:dth4@st-andrews.ac.uk (DTH)}, Simon P. Driver$^{1,2,3,4}$, Ewan Cameron$^{1,2}$, Nicholas Cross$^{1,5}$, \newauthor Jochen Liske$^{6}$ and Aaron Robotham$^{1,2}$\\
$^1$Scottish Universities Physics Alliance (SUPA)\\
$^2$School of Physics \& Astronomy, University of St Andrews, 
North Haugh, St Andrews, Fife, KY16 9SS, UK\\
$^3$Distinguished Visitor at the Australian Telescope National Facility, Marsfield, NSW 2122, Australia\\
$^4$Visiting Professor at the University of Western Australia, School of Physics, Crawley, WA 6009, Australia\\
$^5$Institute of Astronomy, University of Edinburgh, Royal Observatory, Edinburgh, EH9 3HJ, UK\\
$^6$European Southern Observatory, Karl-Schwarzschild-Str. 2, 85748 Garching bei M\"{u}nchen, Germany
}

\date{Received 2009 November 27; Accepted 2010 January 18; in original form 2009 August 17}
\pubyear{2008} \volume{000}
\pagerange{\pageref{firstpage}--\pageref{lastpage}}

\maketitle
\label{firstpage}

\begin{abstract}
We combine data from the MGC, SDSS and UKIDSS LAS surveys to produce $ugrizYJHK$ luminosity functions and densities from within a common, low redshift volume ($z<0.1$, $\sim 71,000 h_1^{-3}$Mpc$^3$ for $L^*$ systems) with 100 per cent spectroscopic completeness.  In the optical the fitted Schechter functions are comparable in shape to those previously reported values but with higher normalisations (typically 0, 30, 20, 15, 5 per cent higher $\phi^*$-values in $u, g, r, i, z$ respectively over those reported by the SDSS team). We attribute these to differences in the redshift ranges probed, incompleteness, and adopted normalisation methods. In the NIR we find significantly different Schechter function parameters (mainly in the $M^*$ values) to those previously reported and attribute this to the improvement in the quality of the imaging data over previous studies. This is the first homogeneous measurement of the extragalactic luminosity density which fully samples both the optical and near-IR regimes. Unlike previous compilations that have noted a discontinuity between the optical and near-IR regimes our homogeneous dataset shows a smooth cosmic spectral energy distribution (CSED). After correcting for dust attenuation we compare our CSED to the expected values based on recent constraints on the cosmic star-formation history and the initial mass function. 
\end{abstract}


\section{Introduction}
The galaxy luminosity distribution at any given epoch is a fundamental observable feature of the universe. It is an account of how the space density of galaxies varies with flux, and provides an insight into how visible matter is fragmented. This can be used to constrain both galaxy evolution \citep{tex:benson} and structure formation models \citep{tex:colemodel}. For over half a century astronomers have attempted to parametrise the galaxy luminosity distribution. The current standard, the Schechter function \citep{tex:schechter}, built upon work by \citet{tex:hubble}, \citet{tex:zwicky} and \citet{tex:abell}, amongst others. \\
Observations in visible light, and in particular the wavelength range $350$---$550$nm, are generally dominated by the most recently formed stellar population, whereas the light in the near infrared (NIR) is typically dominated by the longer lived, lower mass stars that constitute the bulk of the stellar mass (modulo some contamination from the AGB branch). The NIR is also less susceptible to internal dust attenuation \citep{tex:calzetti}, with an estimated $\sim80$ per cent, $\sim50$ per cent and $\sim20$ per cent of the integrated flux from galaxies being attenuated in the UV, optical and NIR respectively (see \citealt{tex:uvopnir}). These two benefits, the focus on stellar mass and lower attenuation, make NIR luminosity functions arguably more useful for characterising the underlying properties of the galaxy population, for comparison with semi-analytical models, and, as a stepping stone to calculating the yet more fundamental stellar mass function (see for example \citealt{tex:bagldri}). However recent results, particularly in the $K$ band, show a relatively large range in the reported Schechter function parameters and do not yet span the full NIR wavelength range available; for example no $Y$ band LF has yet been reported. The first aim of this paper is to provide fresh estimates of the NIR luminosity functions using the latest available data from the UKIRT Infrared Deep Sky Survey Large Area Survey in the $Y, J, H$ \& $K$ passbands.\\
By integrating over the product of luminosity and the luminosity function, one derives another useful measurement: the total luminosity density, $j$ (sometimes written as $\mathcal{L}$) at a specific wavelength. The form this takes for a single component Schechter function in solar luminosity units ($hL_{\astrosun,\lambda}$Mpc$^{-3}$) can be seen in Equation \ref{eqn:totlum}. \\
\begin{equation} \label{eqn:totlum}
j_{\lambda} = \phi^{*}_{\lambda} 10^{-0.4(M^{*}_{\lambda} -
M_{\astrosun,\lambda})} \Gamma(\alpha_{\lambda} + 2)
\end{equation}
where $\phi^*_{\lambda}, M^*_{\lambda}$, and $\alpha_{\lambda}$ are the wavelength dependent Schechter function parameters.  Measuring luminosity densities across the full UV/optical/NIR wavelength range (where starlight entirely dominates the energy output), allows one to build up the cosmic spectral energy distribution (CSED) for a representative volume of the local Universe. This is important as it provides a description of the mean radiation field from UV to NIR in which the contents of the nearby Universe are lying. However the radiation we detect is only some fraction of that produced with a significant amount attenuated by dust within the host galaxies \citep{tex:dustatt}.  Correcting for the dust attenuation (see \citealt{tex:uvopnir}), one can derive the pre-attenuated CSED. We can use this empirical result to test our understanding of the cosmic star-formation history (e.g. \citealt{tex:madau}, \citealt{tex:bagl}, \citealt{tex:wilkins}).\\
 One of the reasons this approach has not taken off is an apparent discontinuity between the optical and NIR luminosity density measurements which is clearly unphysical. This was first reported in \citet{tex:wrighte} where an extrapolation of the SDSS Early Data Release luminosity density measurements in $ugriz$ led to an overprediction of the observed NIR luminosity densities by a factor of 2-3. This issue was mainly resolved when revised SDSS LF estimates were published which incorporated evolutionary effects \citep{tex:blanton}, however the problem has not entirely vanished and is clearly noticeable in \citet{tex:bagl} where data from the SDSS \citep{tex:blanton}, GALEX \citep{tex:galex} and the NIR studies of \citet{tex:cole}, \citet{tex:kochanek} were used to produce the nearby CSED.  This analysis actually brought to light a number of issues including the incomplete coverage of the NIR wavelength range (with only $J$ and $K$ values available), the significant scatter in the available $K$-band data, and an apparently unphysical step-function between the optical and the $J$-band resulting in the $J$-band data being dropped. This discontinuity from the optical to near-IR echoes that seen in \citet{tex:wrighte} and is also obvious when comparing the recent 6dfGS survey measurements of the galaxy luminosity density in $JHK$ \citep{tex:hj} with recent SDSS results in $ugriz$ (\citealt{tex:blanton}; \citealt{tex:sdssdr6}). In general any discrepancy in the CSED between optical and NIR wavelengths could in part be explained by: a very top heavy IMF, cosmic variance in the NIR data (e.g. \citealt{tex:somerville}), surface brightness selection bias in the NIR data \citep{tex:ajsmith}, discrepancies in the photometric calculation, or spectroscopic incompleteness bias. Certainly, when one reviews the most recently published NIR luminosity densities one does find significant scatter. In particular, \citet{tex:kochanek} and \citet{tex:huang} examined insufficiently large volumes to overcome cosmic variance. Other attempts, such as \citet{tex:cole} and \citet{tex:hj}, have probed greater volumes but are dependent on shallow 2MASS imaging data that has been shown to be susceptible to surface brightness (SB) bias, missing both galaxies and flux (e.g. \citealt{tex:andreon}, \citealt{tex:bell}, \citealt{tex:eke} and \citealt{tex:kirby}). \\
The UKIRT Infrared Deep Sky Survey Large Area Survey (UKIDSS LAS) extends $\sim2.7$ mag deg$^{-2}$ deeper than 2MASS in $K$, and therefore should be less susceptible to SB effects. \cite{tex:ajsmith} recently produced a $K$-band SDSS-UKIDSS luminosity function that appears to agree closely with the results of \citet{tex:cole} and \citet{tex:hj}. This would suggest that 2MASS and UKIDSS photometry are consistent and the reported NIR LFs robust. However, due to an unresolved issue with the UKIDSS extraction software affecting $\sim10$ per cent of the UKIDSS data (see detailed discussion in Appendix \ref{subsec:deblend}), \citeauthor{tex:ajsmith} question the validity of their own results. \citeauthor{tex:ajsmith} also restricted their analysis to $K$-band only where, for the purpose of recovering the cosmic SED, it is obviously desirable to recover measurements in all available filters (i.e. $YJHK$). Finally a significant issue may well arise from cosmic variance \citep{tex:drivcosvar} with NIR surveys typically being based on more local samples (e.g., 6dfGS with $\langle z \rangle \approx 0.05$) and the optical data being based on deeper samples (e.g., SDSS with $\langle z \rangle \approx 0.1$). In an ideal situation one would seek to derive the full cosmic SED from within a single survey where the impact of cosmic variance will affect the measurements at all wavelengths in a similar manner (modulo colour dependent clustering).  In this paper we do precisely this; by combining data from the Millennium Galaxy Catalogue (MGC), SDSS and UKIDSS LAS surveys we derive the $ugrizYJHK$ luminosity distributions and pre- and post attenuated cosmic SEDs within a single well understood volume of $\sim 71,000h^3$Mpc$^{3}$.\\
In Section \ref{section:data} we briefly describe the three surveys, their shared area and appropriate flux limits (to ensure against colour bias). In Section \ref{section:LF} we describe our methodology for deriving Schechter luminosity function parameters.  Finally, in Sections \ref{section:discuss} and \ref{section:conclude} we present our results, discuss how they compare with previous work and what conclusions we can make. We also include an outline (Appendix \ref{subsec:deblend}) of the problems with the CASU source extraction in early UKIDSS data releases, and the steps we undertook to reanalyse UKIDSS data for this paper. Throughout we adopt an $h=1, H_{0} = 100 h $ km s$^{-1}$ Mpc$^{-1}$, $\Omega_{M} = 0.3, \Omega_{\Lambda} = 0.7$ cosmological model. All optical magnitudes are quoted in the SDSS photometric system (which is consistent with the AB system $\pm 0.04$mag) and all NIR magnitudes in the UKIDSS preferred Vega system unless otherwise stated (Table. 1 shows the relevant conversions between the AB and Vega systems).

\section{Data} \label{section:data}
In this paper we combine data from three distinct datasets: the Millennium Galaxy Catalogue (MGC; \citealt{tex:mgc}; \citealt{tex:z}), the Sloan Digital Sky Survey (SDSS; \citealt{tex:sdssyork}; \citealt{tex:sdssdr5}), and the UK Infrared Deep Sky Survey Large Area Survey (UKIDSS LAS; \citealt{tex:ukirt}; \citealt{tex:ukidsswarren}). The MGC provides the deepest ($B$ $<20$ mag) wide area redshift survey to date (extending about 1.5 mag fainter than SDSS main), while the SDSS provides full optical coverage ($ugriz$), and the UKIDSS LAS full near-IR coverage ($YJHK$). The combination of these datasets enable the full NIR-optical mapping of the cosmic energy output of the Universe emanating from a representative volume, robust to both wavelength dependent cosmic variance and spectroscopic incompleteness. In this section we briefly introduce each survey and describe the coverage and flux limits of the common region.

\begin{figure}
\includegraphics[width=220pt]{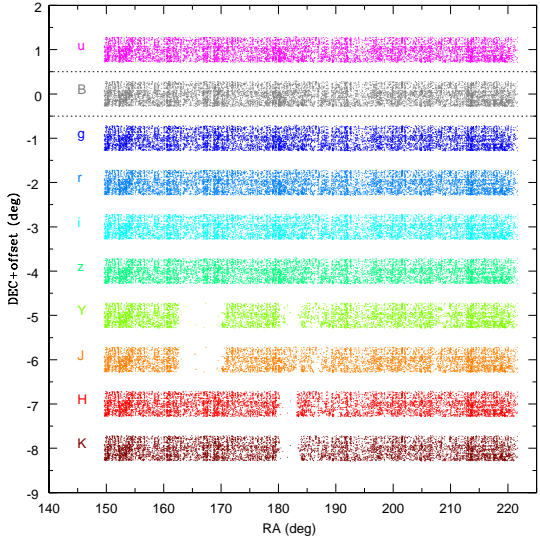}
\caption{The distribution of galaxies along the common MGC-SDSS-UKIDSS region for (top-to-bottom) the $uBgrizYJHK$ bands (as indicated). Note that the MGC data does not provide contiguous coverage within the designated area due to gaps between CCDs (the "L" shaped layout of the CCDs causes this problem) and masked regions around bright stars and satellite trails. The SDSS DR5 coverage of the MGC is complete while the UKIDSS LAS shows some clear gaps. The coverage as a fraction of the total area is summarised in Column 2 of Table \ref{coverage}.}
\label{fig:grid}
\end{figure}

\subsection{The Millennium Galaxy Catalogue (MGC)} \label{section:MGC}
The Millennium Galaxy Catalogue \citep{tex:mgc} is a deep ($B_{\rm lim}=24$ mag, $\mu_{\rm lim}$ = 26 mag arcsec$^{-2}$), $B$ band galaxy survey created using the Wide Field Camera on the 2.5m Isaac Newton Telescope, with observations taken between 1999 and 2001. Surveying a long (75$^{\circ}$), thin (0.5$^{\circ}$), equatorial strip amounting to 37 deg$^{\rm 2}$ of sky (30.88 deg$^{\rm 2}$ after cleaning and cropping), the MGC contains data on over a million objects, with 10,095 galaxies brighter than $B = 20$ mag (this resolved sub-catalogue is referred to as MGC-Bright, and the integrity of every object within it has been verified by eye and fixed where necessary). Using the extensive MGC overlap regions, for objects in the range $17 \leq B_{\rm MGC} \leq 21$ mag, the astrometric rms uncertainty has been shown to be $\pm$0.08 arcsec, and the internal photometric uncertainty $\pm$0.023 mag. The seeing ranged from 0.9 to 2 arcsec, with a median of 1.3 arcsec. Object detection was achieved using the SExtractor program (see \citealt{tex:sext}), with a constant surface brightness threshold, $\mu_{lim}$, of 26 mag arcsec$^{-2}$, and galactic extinction corrected using dust maps created by \cite{tex:schl}. \citet{tex:z}, have obtained redshifts for 96 per cent of the MGC-Bright galaxies using the SDSS and 2dFGRS data releases combined with a dedicated MGCz 2dF survey. \citet{tex:allen} have conducted bulge disc decomposition using GIM2D for the entire MGC-Bright catalogue. All data are publicly available from http://www.eso.org/$\sim$jliske/mgc/.

\subsection{The Sloan Digital Sky Survey} \label{section:SDSS}
The Sloan Digital Sky Survey (SDSS, \citealt{tex:sdssyork}) is the largest combined photometric and spectroscopic survey ever undertaken, containing spectra of $\sim$930k galaxies and imaging of over $11500$ deg$^{\rm 2}$ of sky, using five filters with average wavelengths between $300$ and $1000$ nm ($ugriz$). SDSS data has been publicly released in a series of 7 data releases. We are using the fifth data release catalogue (SDSS-DR5), which covers approximately $8000$ deg$^{\rm 2}$ and contains spectroscopy for 675k galaxies. The MGC region falls within the SDSS DR5 area of coverage.\\
Using the SDSS DR5 database stored in the WFCAM Science Archive\footnote{The WSA (\citealt{tex:wfcam}) is a storage facility that contains copies of catalogues from a number of surveys, including the SDSS and the MGC, as well as being the primary store for raw UKIDSS data.}, we output a complete list of SDSS PhotoPrimary objects within the MGC region. We then matched MGC-Bright objects to SDSS PhotoPrimary objects using the STILTS catalogue matching tool \citep{tex:stilts}, with a maximum centroid separation tolerance of 2.5 arcsec. Where there are multiple matches within 2.5 arcsec to one MGC object, STILTS takes the closest matching object. For the purposes of photometry we adopt the extinction corrected SDSS Petrosian apparent magnitudes ($\text{PetroMag}_X-\text{extinction}_X$, where $X$ is $u$, $g$, $r$, $i$ or $z$). Our final DR5-MGC-Bright matched catalogue contains 10050 SDSS-MGC matching galaxies (99.6 per cent of the 10095 sources that make up the MGC-Bright catalogue). The coverage of the MGC by SDSS DR5 is shown in Fig.~\ref{fig:grid}. SDSS fluxes are reported throughout in AB magnitudes, conversions to Vega are shown in Table.\ref{tab:offsets}.

\subsection{UKIRT Infrared Deep Sky Survey (UKIDSS)} \label{section:UKIDSS}
UKIDSS \citep{tex:ukirt} is a seven year near-infrared survey programme that will cover several thousand deg$^{\rm 2}$ of sky. The programme utilises the Wide Field Camera (WFCAM) on the 3.8m United Kingdom Infra-Red Telescope (UKIRT). The full UKIDSS program consists of five separate surveys, each probing to a different depth and for a different scientific purpose. The shallowest of these surveys, the UKIDSS Large Area Survey (LAS), contains the full MGC region.\\
When complete, the LAS will cover 4000 deg$^{\rm 2}$ of sky to target depths ($5\sigma$ point source detections in Vega) of $K=18.2$ mag, $H=18.6$ mag, $J=19.9$ mag (after two passes; this paper uses only the first $J$ pass which is complete to 19.5 mag) and $Y=20.3$ mag (for conversion to the AB system please see the offsets shown in Table.~\ref{tab:offsets}). UKIDSS data are required to have a seeing FWHM of $<1.2$ arcsec, photometric rms uncertainty of $<0.02$ mag and astrometric rms of $<0.1$ arcsec. Each position on the sky is targeted for 40s per pass. All survey data for this paper is taken from the third data release (DR3PLUS).\\
During the course of our analysis a number of problems were encountered with the online catalogues generated by the Cambridge Astronomical Survey Unit (CASU), particularly the calculation of Petrosian fluxes for deblended systems (as first noted in \citealt{tex:ajsmith}). The problems are described in detail in Appendix \ref{subsec:deblend} and were considered sufficiently insurmountable to warrant re-deriving the catalogues in entirety from the reduced data. The derivation of our final MGC-Y,J,H and K catalogues using SExtractor (\citealt{tex:sext}) are also described in Appendix \ref{subsec:deblend}.\\
In summary after revision of the UKIDSS photometry we estimate that individual fluxes in the $Y, J, H,$ and $K$ bands are credible to $\pm 0.05$ mag. This value derives from a comparison between the original UKIDSS LAS data for which deblending was not required, and our revised catalogue.

\begin{table}
\caption{The Absolute magnitude of the Sun in various filters for the AB and Vega systems along with the approximate filter central wavelength. These values were derived for us by Paul Hewett (priv. comm). $M_{\astrosun}$ differ from those in Table 1 of \citet{tex:blantonroweis} by $0$ mag, $0.03$ mag, $0.07$ mag, $0.03$ mag and $0.03$ mag in the $u$, $g$, $r$, $i$ and $z$ bands.}
\label{tab:offsets}
\begin{tabular}{p{0.1cm}cccc} \hline \hline
& $\lambda$ ($\mu$m) & $M_{\astrosun} (AB)$ & $M_{\astrosun} (Vega)$ & $ M_{\astrosun} (AB) - M_{\astrosun} (Vega)$\\ \hline
u&0.3546 & 6.38&5.47&0.91\\
g&0.4670 & 5.15&5.23&-0.08\\
r&0.6156 & 4.71&4.55&0.16\\
i&0.7471 & 4.56&4.19&0.37\\
z&0.8918 & 4.54&4.00&0.54\\
Y&1.0305 & 4.52&3.89&0.63\\
J&1.2483 & 4.57&3.63&0.94\\
H&1.6313 & 4.71&3.33&1.38\\
K&2.2010 & 5.19&3.29&1.90\\
\hline
\end{tabular}
\end{table}

\setcounter{figure}{1}
\begin{figure*}
\includegraphics[width=440pt]{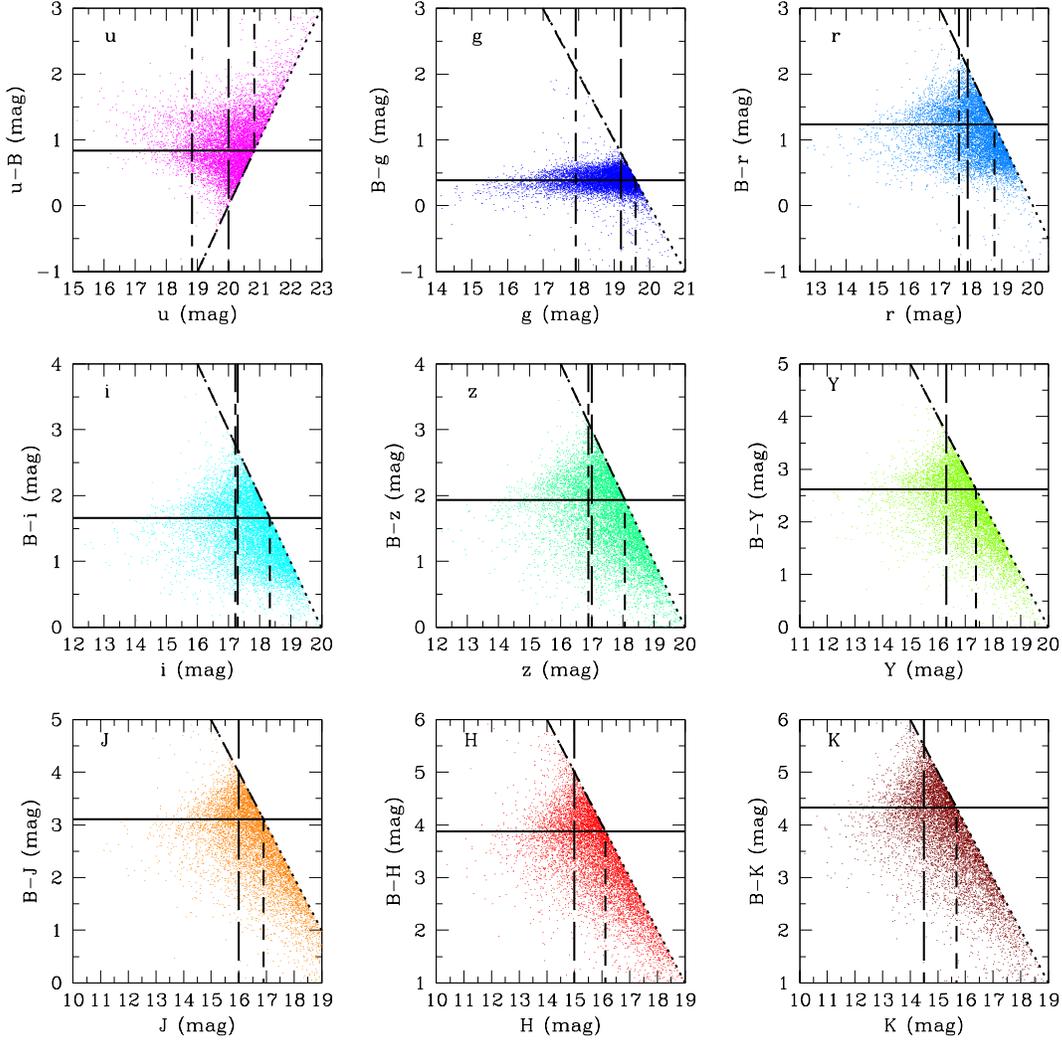}
\caption{The colour-magnitude diagrams for the $ugrizYJHK$ filters versus $B$. The vertical lines are our apparent magnitude limits in $ugrizYJH$ or $K$, the diagonal lines are our $B$ band magnitude limits.}
\label{fig:mcol}
\end{figure*}

\subsection{The MGC-SDSS-UKIDSS LAS common area}
Figure~\ref{fig:grid} shows the coverage of the MGC region by the SDSS DR5 and UKIDSS LAS datasets. Note that the MGC footprint itself does not provide continuous coverage within a rectangle but rather a square-tooth profile (the INT WFC has a thick ``L''-shaped footprint). The MGC region has also been carefully masked to remove objects close to bright stars, where the flux might be compromised, satellite trails, CCD defects, and edge effects within the MGC region which reduces the effective area on sky from $\sim$38 deg$^{\rm 2}$ to 30.88 deg$^{\rm 2}$. This process negates any issue of the blocking factor of distant galaxies by bright foreground stars as well as false detections due to spurious light scattering (e.g. diffraction spikes and ghosting). Full details of the MGC footprint and its masking is given in \citet{tex:mgc}.\\
SDSS DR5 provides complete coverage of the cleaned MGC region in all bands (see Fig.~\ref{fig:grid}, $ugriz$). A detailed comparison of the photometry, astrometry, and deblending between the MGC and DR1 was described in detail in \citet{tex:crosssdss}. In total MGC-Bright contains 10,095 galaxies to $B_{MGC}=20.0$ mag of which matching detections are identified in the SDSS survey for 99.6 per cent of the MGC targets, the majority of failed matches comes from extreme low surface brightness systems and differences in deblending decisions.\\
The UKIDSS LAS DR3PLUS does not have complete coverage of the combined MGC-SDSS region (Fig.~\ref{fig:grid}, $YJHK$) and the level of completeness varies for each filter. The gaps are due to data failing the UKIDSS data control process (typically seeing and sensitivity critereon). From the common regions the distribution of galaxies that have been flagged as containing major errors or requiring deblending (see Appendix \ref{subsec:deblend}), show no obvious bias along the strip (as one would expect given the quality control process). The area of overlap can therefore be derived based on the fraction of MGC galaxies with UKIDSS data available. The coverage of the SDSS and UKIDSS-LAS datasets is summarised in Columns 2 and 3 of Table \ref{coverage}.\\
The MGC is taken as the master catalogue in all that follows as it is the deepest (in terms of flux sensitivity), contains the highest signal-to-noise detections and has been fully masked and eyeballed. Also, where necessary, all objects within it have already been reconstructed or deblended.\\
\setcounter{table}{1}
\begin{table*}
\caption{Parameters defining the coverage and depth(s) of the joint MGC-SDSS-UKIDSS common region along with the adopted $K(z)$ corrections and $E(z)$ ranges. Though we set $E(z)=0$, we use the $\beta$ ranges in column 8 to calculate the scale of the uncertainty due to evolution, and the range of $K(z)$-corrections in column 7 to calculate the uncertainty due to the $K$-correction. The sample size column gives the number of galaxies brighter than Limit 2 within the defined redshift limits. The $\Delta m_{z=0.1}$ column gives the range of effect the combined K+E correction can have on a $z=0.1$ galaxy.} \label{coverage}
\begin{tabular}{cp{1.4cm}p{1.05cm}p{1cm}p{1cm}p{1.8cm}ccc} \hline \hline
Filter & Coverage (per cent) & Area (deg$^{\rm 2}$) & Limit 1 (mag) & Limit 2 (mag) &Sample size (0.0033$<$$z$$<$0.1)& $k(z)$ & $E(z)$ $\beta$ range & $\Delta m_{z=0.1}$\\ \hline
$u$ & $100\%$ & 30.88 & 20.0 & 20.84 & 3267 & $(2.16^{+0.95} _{-0.87})z$ & -1.36---0 & -0.01---0.31\\
$g$ & $100\%$ & 30.88 & 19.2 & 19.61 & 3328 & $(2.71^{+0.74} _{-1.09})z$  & -0.68---0 & 0.09---0.35\\ 
$r$ & $100\%$ & 30.88 & 17.9 & 18.76 & 2781 & $(0.95^{+0.34} _{-0.52})z$  & -0.45---0 & 0.00---0.13\\ 
$i$ & $100\%$ & 30.88 & 17.3 & 18.34 & 2623 & $(0.48^{+0.38} _{-0.28})z$   & -0.34---0 & -0.01---0.09\\ 
$z$ & $100\%$ & 30.88 & 17.0 & 18.07 & 2437 & $(0.03^{+0.33} _{-0.18})z$    & -0.27---0 & -0.01---0.04\\
$Y$ &  $77\%$ & 23.69 & 16.3 & 17.38 & 1798 & $(0.00^{+0.24} _{-0.12})z$    & -0.23---0 & -0.04---0.02\\
$J$ &  $81\%$ & 25.07 & 16.0 & 16.89 & 1589 & $(-0.61^{+0.27} _{-0.10})z$   & -0.19---0 & -0.08--- -0.03\\
$H$ &  $91\%$ & 27.89 & 15.0 & 16.12 & 1890 & $(-0.28^{+0.24} _{-0.12})z$   & -0.17---0 & -0.06---0.00\\ 
$K$ &  $91\%$ & 27.99 & 14.5 & 15.67 & 1785 & $(-1.44^{+0.10} _{-0.02})z$  & -0.15---0 & -0.16--- -0.13 \\ \hline
\end{tabular}
\end{table*}

\subsection{$ugrizYJHK$ magnitude limits}\label{subsec:apmlimit}
\begin{figure}
\includegraphics[width=220pt]{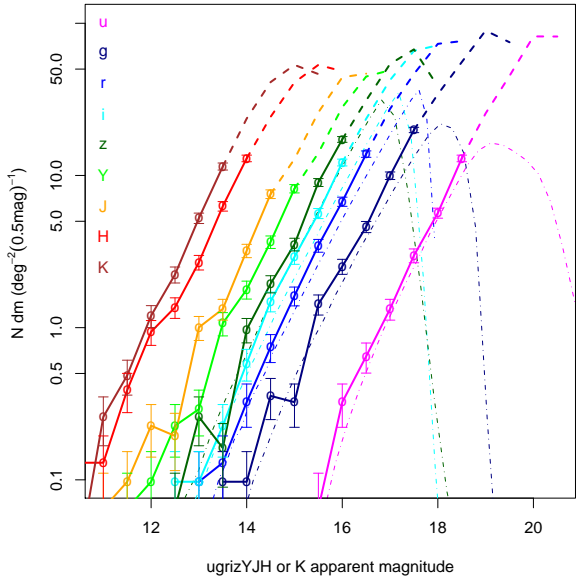}
\caption{The number of galaxies, by apparent-magnitude in the $ugrizYJHK$ filters. The dash-dot-dash lines are galaxy counts from \citet{tex:sdssdr6}. \citeauthor{tex:sdssdr6} justify their early downturn as an issue of redshift incompleteness, and introduce conservative magnitude cuts accordingly.}
\label{fig:numc}
\end{figure}

\begin{figure}
\includegraphics[width=220pt]{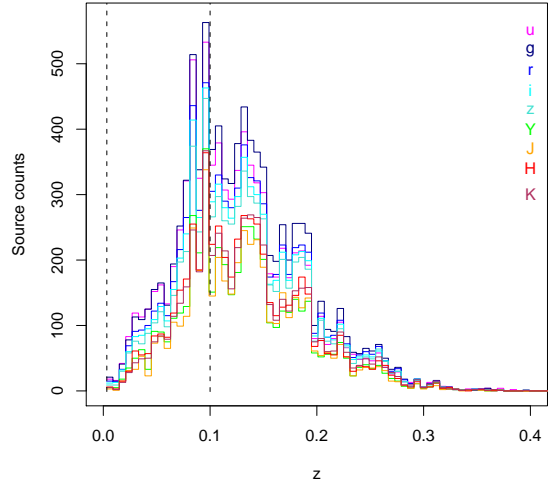}
\caption{The redshift distribution of galaxies within our apparent-magnitude limited matched samples, in bins of z=0.006. The dashed lines signify our sample redshift limits.}
\label{fig:zdist}
\end{figure}

In order to derive luminosity distributions for multi-wavelength data from a $B$ band spectroscopic sample it is important to consider the colour bias. Figure~\ref{fig:mcol} shows the $B_{MGC}-X$ v $X$ colour plots where $X$ denotes $ugrizYJH$ or $K$. The most conservative approach to deriving an unbiased luminosity distribution is to simply define a complete sample within each band, i.e. cut the sample at a sufficiently bright flux where the full colour distribution is fully sampled (see Fig.~\ref{fig:mcol}, long dashed line) and where the number-counts have yet to show any indication of a turn-down (Fig.~\ref{fig:numc}). Fig.~\ref{fig:numc} also shows that we can extend our samples deeper than those used in \citet{tex:sdssdr6}; our turn-down occurs roughly one magnitude deeper in each SDSS passband. We show the \citeauthor{tex:sdssdr6} limits on Fig.~\ref{fig:mcol} as long dash-short dash lines. In all SDSS passbands they are more conservatively cut than our most conservative limit, particularly in the $u$ band. The most liberal approach is to use all the data and define a unique flux limit for each individual object based upon the spectroscopic limit of $B_{\text{MGC}}=20.0$ mag combined with the object's colour, i.e. $X_{\text{limit}}=20.0-(B-X)$, using these $B$ and $X$ limits to appropriately weight each object (Fig.~\ref{fig:mcol}, dotted line). While the former will reduce the sample size significantly the latter will incorporate a large quantity of data in the regime where flux measurements may not be credible. Here we adopt a hybrid approach (Fig.~\ref{fig:mcol}, short dashed line) where we determine the mean colour for each sample (solid line) and combine this with the spectroscopic limit of $B_{\text{MGC}}=20.0$ mag to determine a nominal limit, i.e. the limit where the mid-point of the colour distribution is spectroscopically sampled.  Each galaxy is then allocated a flux limit which is the brighter of the nominal limit or that defined by the locus $X_{\text{limit}}=20.0-(B_{\text{MGC}}-X)$, see short dashed lines on Fig.~\ref{fig:mcol}. The conservative limits (Limit 1), nominal limits (Limit 2), and effective sample sizes are shown in Table~\ref{coverage}. Table~\ref{colour} shows the median colour and 3$\sigma$-clipped standard deviation derived from the data above the conservative limit (Limit 1). After imposing our magnitude limits, every galaxy within our samples has a redshift. Figure \ref{fig:zdist} shows the distribution of our apparent magnitude cut samples by redshift.\\

\begin{table}
\caption{Median colours and $3\sigma$ clipped standard deviations above the completeness limits defined by Flux limit 1 in Table.~\ref{coverage}.} \label{colour}
\begin{tabular}{ccc} \hline \hline
Colour & Median & Std (3$\sigma$ clipped) \\ \hline
$u-B$ & 0.81 & 0.40 \\
$B-g$ & 0.38 & 0.13 \\
$B-r$ & 1.21 & 0.44 \\
$B-i$ & 1.62 & 0.57 \\
$B-z$ & 1.94 & 0.59 \\
$B-Y$ & 2.56 & 0.62 \\
$B-J$ & 3.06 & 0.51 \\
$B-H$ & 3.82 & 0.59 \\
$B-K$ & 4.25 & 0.59 \\ \hline
\end{tabular}
\end{table}

\section{Methods} \label{section:LF}
\subsection{Luminosity distribution and function measurement} 

\subsubsection{The Schechter function} \label{subsec:glf}
The commonly adopted functional form of the luminosity function is the Schechter luminosity function \citep{tex:schechter}:
\begin{equation} \label{eqn:schech}
\frac{dn}{dM} = \phi(M) = 0.4\text{ } \rm{ln 10}\text{ } \phi^{*} \frac{(10^{0.4(M^{*}-M)} )^{\alpha+1}}{e^{10^{0.4(M^{*}-M)}}} 
\end{equation}
Where $M^{*}$ is the characteristic magnitude above which the power law behaviour is suppressed, with the power law having a slope of $\alpha + 1$. The $\phi^{*}$ parameter is a normalisation constant. Whilst only an approximate match to the galaxy luminosity distribution\footnote{\citet{tex:hj} found that it does not turn down sharply enough at $M^{*}$, \citet{tex:blanton} found that there is a deviation from the Schechter function at the luminous end and a strong positive correlation between the $M^{*}$ and $\alpha$ parameters, and \citet{tex:bell} have attempted to modify it to fit their dataset more accurately.}, it is the standard parametric fit adopted in the literature and therefore useful for comparing with earlier work.

\subsubsection{SWML with variable magnitude limits}
There are a number of different techniques for measuring the galaxy luminosity distribution (see the comparison by \citealt{tex:willmer}). We adopt the SWML method, presented and described in detail in \citet{tex:eep}. This is a maximal-likelihood method of calculating volume corrected binned luminosity distributions in a non-parametric manner. The resulting luminosity distribution can then be fit via $\chi^2$-minimisation to determine the Schechter function parameters. As noted, the SWML method makes no a priori assumption of the form of the luminosity function, but does require an additional method for normalisation. Here this is done by (1) selecting an absolute magnitude range ($M_1$ to $M_2$), (2) finding the number of galaxies inside this range ($N_{1 \rightarrow 2}$), (3) calculating the maximum and minimum redshifts over which galaxies of magnitude $M_1$ to $M_2$ can be seen and (4) integrating the standard expression for the volume interval ($V_{1 \rightarrow 2} = \frac{\delta V}{\delta z}$) over this redshift range for the specified cosmology. The SWML luminosity distributions are then scaled such that: $\phi($M$_1$ to $M_2$) = $\frac{N_{1 \rightarrow 2}}{V_{1 \rightarrow 2}}$. The normalisation is therefore a $1/V_{\text{max}}$-method with the SWML luminosity distribution rescaled to produce the required number of galaxies within the specified absolute magnitude range. Care must be taken to ensure that the calibration volume is complete for the range of absolute magnitudes selected. \citet{tex:willmer} has found that, in samples where the faint end of the luminosity distribution is under-represented, the behaviour of the SWML method can be eccentric, with a larger $\alpha$ than expected being recovered. In \citet{tex:eep} the method adopts a constant flux limit for all galaxies. We introduce a minor modification to accommodate for the non-uniform flux limit (because of the colour bias) with our final statistic given in Equation \ref{eqn:swml}, as:
\begin{equation}
\begin{gathered}
\phi(M_{k}) dM = \frac{ \sum_{i=1}^{N_{g}} W(M_{i} - M_{k}) }{ \sum_{i=1}^{N_{g}} \frac{ H(M_{k} - M_{\text{faint}(z_{i},m_{\text{lim},i})}) }{\sum_{j=1}^{N_{g}} \phi_{M_{j}} dM H(M_{j}-M_{\text{faint}(z_{i},m_{\text{lim},i})})} } \\
\text{where:}\\ 
W(x) = \begin{cases} 
1 & \text{if } -\frac{dM}{2} \le x \le \frac{dM}{2} \\
0 & \text{otherwise}
\end{cases} \\
H(x) = \begin{cases}
1 & \text{if } x \le -\frac{dM}{2} \\
\frac{1}{2} - \frac{x}{dM} & \text{if } -\frac{dM}{2} \le x \le \frac{dM}{2} \\
0 & \text{otherwise}
\end{cases}
\end{gathered}
\label{eqn:swml}
\end{equation}
The key distinction here is the introduction of an individual magnitude limit ($m_{\text{lim},i}$) for each object. This modification is identical to that adopted by the 2dFGRS, 6dfGS and MGC teams when dealing with non-uniform flux limits.

\subsubsection{Calculation of absolute magnitudes}
As the SWML method in Equation \ref{eqn:swml} requires absolute magnitudes it is first necessary to convert from apparent to absolute magnitudes. This requires a cosmological model and $K(z)$ and $E(z)$-corrections. The adopted cosmological parameters are $\Omega_M$=0.3, $\Omega_{\Lambda}$=0.7, $H_o$=100 km s$^{-1}$ Mpc$^{-1}$, compatible with both \citet{tex:hj} and \citet{tex:blanton}. We reject galaxies with redshifts outside the interval $0.0033 \le z \le 0.10$. The lower limit is defined as that required to overcome the local velocity field (see \citealt{tex:z}), the upper limit is chosen to minimise the uncertainty inherent in the adopted $K(z)$ and $E(z)$ corrections and ensure a uniform survey volume across all wavelengths. For both corrections we elect to adopt global rather than individual values and Monte-Carlo over a suitably broad range of uncertainty to ensure the final uncertainties on the Schechter function parameters is realistic. Unlike the \cite{tex:blanton} and \cite{tex:sdssdr6} papers, we calculate our absolute magnitudes with the SDSS passbands at z=0, rather than at z=0.1.\\
We calculate the $K(z)$-corrections required for our data (column 7 of Table \ref{coverage}) via the 13.2Gyr Sa-type galaxy spectra from the synthetic library of \citet{tex:pogg}, and the range of possible $K(z)$-corrections from the 7.2 Gyr Sa-type (the lower limit of column 7) and 15.0 Gyr El-type galaxy spectra (the upper limit).\\
We do not believe evolution occurs within $0.0033 < z < 0.1$ (the first four black points in Figure 11 of \citealt{tex:prescott} show no evidence of any occurring); we note that its effects would predominantly impact the $M^{*}$ parameter, with only a small change in the $\alpha$ and $\phi^{*}$ parameters. For instance, using the evolution in the $u$ band luminosity density presented in \citet{tex:prescott} ($\beta=-1.36$ using our sign convention) and the prescription of \citet{tex:phillippsdriver} given by Equation \ref{eqn:phil}\footnote{The $\beta$ parameter in Equation \ref{eqn:phil} and the $\beta$ variable in \cite{tex:prescott} will, by definition, have opposing signs.}:
\begin{equation}
E(z)=2.5\beta\log_{10}(1+z)
\label{eqn:phil}
\end{equation}
where $z$ is the redshift of the corrected galaxy, the inclusion of an evolutionary correction modifies the best-fitting $M^{*}$ parameter by $0.09$ mag, the best-fitting $\alpha$ parameter by 0.02 and the best-fitting $\phi^{*}$ parameter by $0.0003$ h$^{\rm 3}$ Mpc$^{\rm -3}$. Our quoted results do not employ an $E(z)$-correction; however we calculate the effect that including one would produce. We calculate the uncertainty due to evolution by deriving the best-fitting Schechter parameters for five equally spaced $\beta$ values from the range shown in column 8 of Table \ref{coverage}, and take the standard deviation in these results as our evolutionary uncertainty. The effects of the $K(z)+E(z)$ corrections are, unsurprisingly, strongest in UV ($\pm0.05$ mag in $M^{*}$ in the $u$ band), and limited in the NIR ($\pm 0.01$ mag in $M^{*}$ in the $K$ band). \citet{tex:blanton} have used the evolution correction as a reason for a flattening of the faint-end of the luminosity function; within our small redshift range we do not believe that this is the case.\\
Finally, luminosity uncertainty can move a galaxy into the wrong absolute-magnitude bin, and this effect increases the counts within the brighter bins by a greater fraction than the fainter bins; i.e. a classical Malmquist bias. Where there is a large uncertainty in luminosity this must be compensated for. However, the typical luminosity uncertainty in our data ($0.04$ mag in $K$ - see section \ref{subsec:seanalysis}, 0.05 mag in $H$, 0.05 mag in $J$ and 0.04 mag in $Y$) is small enough to make this unnecessary.\\

\subsubsection{Normalisation} \label{sec:normal}

Within $30.88$ deg$^{\rm 2}$ of sky, we calculate the normalisation volume over the redshift range ($0.023 < z < 0.097$) to be $71069h^{-3}$Mpc$^{3}$. The same redshift range is adopted for all nine filters to insure against cosmic variance. The redshift range is selected to be complete for a 1 mag range around the $M^*$ point for each of the nine filters. We calculate the scale of the global overdensity using SDSS-DR7 spectoscopic catalogue data for a 5150 deg$^{\rm 2}$ rectangular region of sky (130 to 236 deg RA, 0 to 58 deg Dec, within our redshift range). We find an average source density of 13.00 de-reddened, $M^{*}_{r} \pm 0.5$ mag galaxies per deg$^{\rm 2}$ within this region. In the MGC area, we find 15.90 de-reddened, $M^{*}_{r} \pm 0.5$ mag galaxies per deg$^{\rm 2}$. We assume that the area lost due to bright star holes within the SDSS region is the same fraction as that lost in the MGC region ($\sim$3\%). When this is taken into consideration the MGC is 19.3\% overdense. This overdensity is visible in Figure~\ref{fig:numc}. For each filter we modify this volume to take into account the variation in the area of coverage, spectroscopic incompleteness (for all of our samples this = 1), and the global over-density of the MGC region, i.e.:
\begin{equation}
V_X=V_B. f_{A,X} .f_{C,X}/f_{\text{MGC}}
\end{equation}
Where $X$ is the filter, $f_{A,X}$ is the fraction of the MGC covered by filter $X$, $f_{C,X}$ is the sample completeness in filter $X$, and $f_{\text{MGC}}$ is the global over-density of the MGC. Using these corrected volumes, and the number of galaxies within a one magnitude range that contains the $M^{*}$ mag galaxies we calculate the galaxy number-density and scale the unnormalised number-densities ($\phi$-values) derived from the SWML method to reproduce this object density:
\begin{equation}
\phi(M)=N_X(M^{*}-0.5 < M < M^{*}+0.5)/V_X
\end{equation}

\section{Results and Discussion} \label{section:discuss}
\setcounter{table}{3}
\begin{table*}
\caption{Derived Schechter function parameters in $ugrizYJHK$ for the magnitude limits indicated within the redshift range $0.0033 < z < 0.1$. The errors shown for the Schechter parameters are, in order, due to the sample size, K(z)-correction and E(z)-correction uncertainties. The errors shown for luminosity density and $\nu f_{\nu}$ statistics are due to sample size, and combined K+E correction uncertainties. These can be combined in quadrature to give the combined error. $u$ and $z$ results have been modified by $-0.04$ mag and $0.02$ mag to compensate for the discrepancy between SDSS and AB magnitude systems.} \label{tab:magcutchanges}
\begin{tabular}{p{1cm}p{3.4cm}p{2.9cm}p{2.7cm}p{1.6cm}p{1.7cm}p{2cm}} \hline \hline
Sample & \textbf{$\phi^{*}$} (h$^{\rm 3}$ Mpc$^{\rm -3}$)& \textbf{$M^{*}$ }(mag) & \textbf{$\alpha$}& \textbf{j} ($\times 10^{8}$ $h$ $L_{\astrosun}$ $\rm{Mpc}^{-3}$) & $\nu f_{\nu}$ ($\times 10^{34}$ $h$ $W$ $Mpc^{-3}$) & $\nu f_{\nu}$ (corrected, same units)\\
\hline
$u<$20.84 & 0.0279$^{+0.0015+0.0004+0.0003} _{-0.0016-0.0004-0.0003}$ &-18.21$^{+0.05+0.04+0.03} _{-0.05-0.04-0.03}$ &-0.93$^{+0.03+0.01+0.01} _{-0.04-0.02-0.01}$&1.91$^{+0.24+0.15} _{-0.25-0.14}$&1.98$^{+0.25+0.15} _{-0.22-0.14}$ &4.50$^{+0.57+0.35} _{-0.50-0.32}$\\
$g<$19.81 & 0.0158$^{+0.0011+0.0002+0.0001} _{-0.0008-0.0002-0.0001}$ &-20.08$^{+0.05+0.06+0.02} _{-0.06-0.06-0.02}$ &-1.15$^{+0.03+0.00+0.00} _{-0.02-0.00-0.00}$&2.17$^{+0.33+0.17} _{-0.34-0.15}$ &5.31$^{+0.80+0.41} _{-0.61-0.38}$ &8.97$^{+1.35+0.68} _{-1.04-0.64}$\\
$r<$18.76 & 0.0124$^{+0.0011+0.0001+0.0002} _{-0.0006-0.0001-0.0002}$ &-20.81$^{+0.08+0.03+0.03} _{-0.05-0.03-0.03}$ &-1.18$^{+0.04+0.01+0.01} _{-0.02-0.00-0.01}$&2.29$^{+0.37+0.14} _{-0.40-0.14}$ &6.35$^{+1.03+0.40} _{-0.93-0.38}$ &9.91$^{+1.60+0.62} _{-1.45-0.58}$\\
$i<$18.34 & 0.0120$^{+0.0010+0.0003+0.0002} _{-0.0008-0.0003-0.0002}$ &-21.16$^{+0.07+0.01+0.01} _{-0.06-0.01+0.01}$ &-1.18$^{+0.03+0.01+0.01} _{-0.03-0.00+0.01}$&2.66$^{+0.47+0.13} _{-0.56-0.12}$ &6.99$^{+1.25+0.33} _{-1.03-0.32}$ &10.3$^{+1.83+0.49} _{-1.52-0.47}$\\
$z<$18.07 & 0.0109$^{+0.0009+0.0002+0.0001} _{-0.0011-0.0002-0.0001}$ &-21.46$^{+0.06+0.01+0.01} _{-0.09-0.01-0.01}$ &-1.18$^{+0.03+0.01+0.00} _{-0.04-0.00-0.00}$&3.07$^{+0.68+0.12} _{-0.47-0.11}$ &6.89$^{+1.53+0.26} _{-1.18-0.25}$ &9.71$^{+2.16+0.37} _{-1.67-0.36}$\\
$Y<$17.38 & 0.0146$^{+0.0021+0.0002+0.0003} _{-0.0019-0.0002-0.0003}$ &-21.94$^{+0.12+0.01+0.01} _{-0.11-0.01-0.01}$ &-1.06$^{+0.08+0.01+0.01} _{-0.07-0.01-0.01}$&3.24$^{+1.08+0.16} _{-0.93-0.15}$ &6.41$^{+2.14+0.31} _{-1.66-0.30}$ &8.66$^{+2.89+0.42} _{-2.24-0.40}$\\
$J<$16.89 & 0.0155$^{+0.0017+0.0002+0.0002} _{-0.0016-0.0002-0.0002}$ &-22.20$^{+0.11+0.01+0.01} _{-0.10-0.01-0.01}$ &-0.90$^{+0.07+0.01+0.01} _{-0.07-0.01-0.01}$&3.17$^{+0.82+0.13} _{-0.72-0.12}$ &4.95$^{+1.28+0.20} _{-1.04-0.19}$ &6.53$^{+1.68+0.26} _{-1.38-0.25}$\\
$H<$16.12 & 0.0149$^{+0.0013+0.0001+0.0001} _{-0.0012-0.0001-0.0001}$ &-23.07$^{+0.08+0.00+0.00} _{-0.08-0.00-0.00}$ &-0.99$^{+0.05+0.00+0.00} _{-0.05-0.00-0.00}$&5.38$^{+1.11+0.05} _{-0.99-0.05}$ &5.65$^{+1.16+0.05} _{-0.95-0.05}$ &6.89$^{+1.42+0.07} _{-1.16-0.07}$\\
$K<$15.67 & 0.0156$^{+0.0015+0.0001+0.0000} _{-0.0014-0.0001-0.0000}$ &-23.36$^{+0.09+0.01+0.00} _{-0.08-0.01-0.00}$ &-0.96$^{+0.06+0.00+0.00} _{-0.05-0.00-0.00}$&6.98$^{+1.49+0.13} _{-1.44-0.13}$&3.48$^{+0.74+0.06} _{-0.65-0.06}$ &4.01$^{+0.85+0.07} _{-0.74-0.07}$\\
\hline
\end{tabular}
\label{tab:sdssmagcutchanges}
\end{table*}

\subsection{$ugrizYJHK$ luminosity distributions, functions and densities} \label{subsec:lumdistfnden}
The resulting luminosity distributions and fitted Schechter functions for all samples ($ugrizYJHK$) are shown in Figure \ref{fig:schfn} and the Schechter parameters tabulated in Table~\ref{tab:magcutchanges}. In general the luminosity functions are reasonable fits to the luminosity distributions based on the reduced-$\chi^2$ values.  The most notable exceptions are in the $z$-band, which appears to show a tentative upturn at fainter magnitudes ($M_z \sim -17$ mag), and the $i$-band, which appears to shows the same effect fainter than $M_i \sim -16.5$ mag. We have checked that the faint-end upturn is not dependent on the faint end magnitude limit; it remains if we cut our samples using our conservative limits (Limit 1 in Table~\ref{coverage}), or if we use the much brighter limits of the SDSS samples. However, we note that the faint-end upturn is confined to these two passbands and does not seem to be a general characteristic of all our luminosity functions. There is no sign of any obvious excess of very bright systems (possibly due to the defined redshift interval as \citeauthor{tex:sdssdr6} found an overdensity at $z>0.1$) and in general the Schechter function provides good fits around the knee and brighter. Overlaid on the diagram are selected recent measurements from other groups. In almost all cases our results lie above those reported from the much larger SDSS survey results. As we are using the SDSS photometry, this cannot be a photometric issue. Furthermore as we have compensated for the MGC's over-density by calibrating to the full SDSS it is also cannot be due to cosmic variance across the sky. A further possibility is simply the difference in adopted $K(z)$ and $E(z)$ corrections as it is noticeable that the vertical offset does appear to show some wavelength dependence with the offset maximum in $g$ ($\sim$30 per cent) and dropping to a minimum in $z$ ($\sim$5 per cent). However we note that the MGC data extends approximately 1 mag deeper in all filters (the SDSS sample limits reported in \citeauthor{tex:sdssdr6} are 19.00, 17.91, 17.77, 17.24, 16.97 mag in $ugriz$ respectively, q.v. Table 2, column 5, and in \citeauthor{tex:blanton} the $ugriz$ limits are 18.36, 17.69, 17.79, 16.91 and 16.50 mag, q.v. Table 1, column 2), our samples have 100 per cent redshift completeness, and we use an identical redshift range for all filters ($0.023 < z < 0.097$). The recent study by \citet{tex:sdssdr6} by comparison has a median redshift completeness of 85 per cent (reported to be both wavelength and flux dependent) that may be partly due to the $\sim$55 arcsec minimum fibre proximity of the SDSS spectral survey and, although having significantly brighter flux limits, is used to probe to significantly higher redshifts ($z \leq 0.2$). Moreover, the normalisation adopted in \citeauthor{tex:sdssdr6} is the \citet{tex:davishuchra} method. This uses the entire data set and tends to overly weight the normalisation towards the higher redshift range where incompleteness may be most severe. Without reanalysing the SDSS data using our methodology it is not possible to ascertain the exact cause of the normalisation discrepancy but it is plausible that it is related to this normalisation method and any bias in the redshift incompleteness.\\
In the NIR the discrepancies are more dramatic, perhaps as expected given the rapid development of NIR technologies. We are not aware of any published $Y$ band luminosity functions for comparison. Once again the offset tends to be that our results produce a higher space-density of galaxies. This is perhaps expected given the significantly deeper imaging data. Comparing our $K$-band result to \citeauthor{tex:ajsmith}, whose data was also based on UKIDSS-LAS, we see excellent agreement at the bright-end but a discrepancy in the faint-end slope.\\
Fig.~\ref{fig:contours} explores the LF shape more closely by showing the $1\sigma$ error contours from our Schechter function fits in the $M^*-\alpha$ plane, thereby illustrating the direction of the known degeneracy between $M^*$ and $\alpha$. Recent values from the literature, as indicated in Table~\ref{tab:othersurveys}, are shown as data points with error bars (colour coded on Fig.~\ref{fig:contours} according to filter). The dashed lines on Fig.~\ref{fig:contours} are the $ugriz$ contours produced when our samples are conservatively cut at the brighter limits used by \citeauthor{tex:sdssdr6}. In general the optical data agree reasonably well with recent results from the two much larger SDSS studies of \citet{tex:blanton} and \citet{tex:sdssdr6}, and the conservatively cut sample and our standard sample 1 $\sigma$ $M^*-\alpha$ contours overlap (except in the $u$ band, which loses the largest fraction of galaxies following the brighter apparent magnitude cut). The volume of the common MGC-SDSS-UKIDSS region is much smaller, and the resulting uncertainty is significantly larger than these two previous SDSS results. As an aside, we note that the two SDSS results, while on the whole are consistent with our results, appear to be inconsistent with each other at a high level of significance. After estimating the $E(Z)$-uncertainty, it seems unlikely that this is the only cause of discrepancy between the two SDSS results (\citeauthor{tex:sdssdr6} do not use an evolution correction, \citeauthor{tex:blanton} does). This suggests that significant unquantified systematics still remain and that the increase in statistical size from the MGC to the entire SDSS DR5 dataset is not necessarily increasing the accuracy to which the measured LFs are known. \\
\setcounter{figure}{4}
\begin{figure*}
\includegraphics[width=440pt]{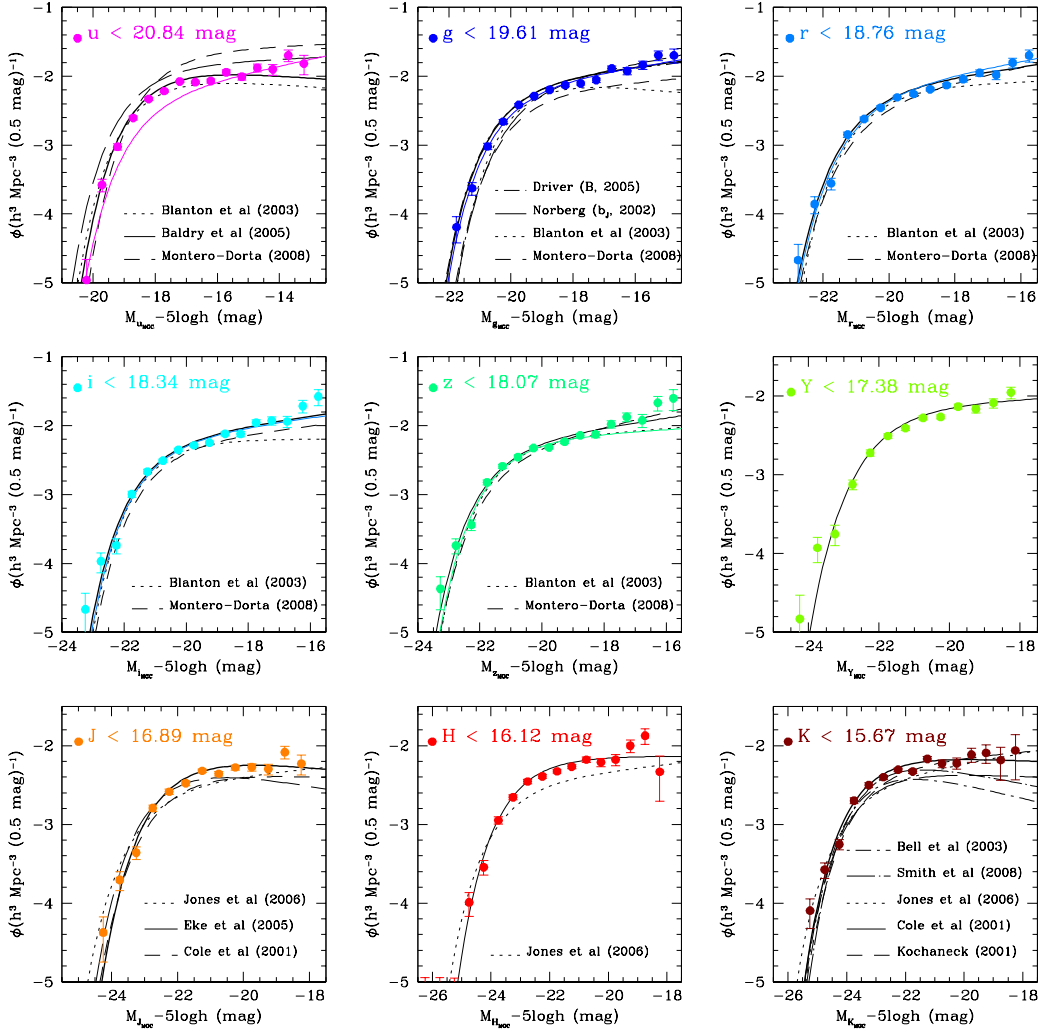}
\caption{$ugrizYJHK$ luminosity distributions and fitted Schechter functions, with comparison lines for Schechter parameters from equivalent surveys. The coloured lines show the best fit Schechter function for $ugriz$ samples that have undergone the more conservative cuts introduced by \citeauthor{tex:sdssdr6}. Poissonean uncertainties are shown for each bin. It should also be noted that the absolute magnitudes in \citet{tex:baldrysdssu}, \citet{tex:blanton} and \citet{tex:sdssdr6} use SDSS passbands that have been redshifted to $z=0.1$, and therefore have been k-corrected (and evolved, where applicable) back to $z=0$. The \citet{tex:pederapm} and \citet{tex:z} comparison lines in the $g$ band use the similar $b_{J}$ and $B$ filters, respectively, which have been transformed to the $g$ band using the assumption that $B-V=0.94$ from \citet{tex:mgc}, and filter conversions in \citet{tex:mgc} and \citet{tex:blantonroweis}.}
\label{fig:schfn}
\end{figure*}
The near-IR LF shapes, in comparison to previous studies, show significant offsets from the MGC-SDSS-UKIDSS data. This presumably reflects the quality of the underlying imaging data with the near-IR data rapidly evolving, in resolution and depth, from the very shallow 2MASS data through to the less shallow UKIDSS LAS. Unfortunately our sample of a few thousand galaxies is insufficient in size to enable a full bi-variate brightness analysis (e.g., \citealt{tex:z}). However, \citet{tex:ajsmith} have quantified the surface brightness limitations of the UKIDSS LAS data in the $K$-band. As the source data for our study is the same as that used by \citet{tex:ajsmith} it is reasonable to adopt their $K$-band limit for completeness of $M_K - 5\log_{10} h=-17.5$ mag. Using the mean colours as indicated in Table.~\ref{colour} we re-derive the $YJHK$ LFs within the revised ranges with no significant change to the Schechter paramaters. This implies that whilst surface brightness selection bias is a concern at some level it is unlikely to be affecting our fits which are dominated by systems at the bright-end of the LF.\\
\begin{figure}
\includegraphics[width=220pt]{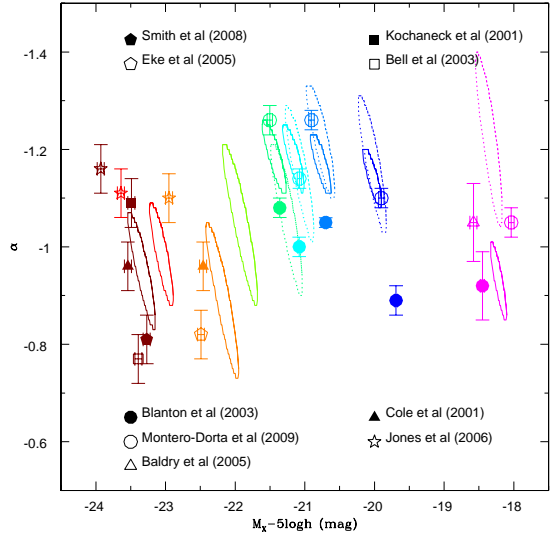}
\caption{$ugrizYJHK$ 1$\sigma$ error contours for the LF fits from Fig.~\ref{fig:schfn}, with LF fits and $\alpha$ uncertainties from equivalent surveys. Dashed contours are from samples that have undergone the more conservative cuts introduced by \citeauthor{tex:sdssdr6}. SDSS equivalent survey points have been transformed from the filter system at $z=0.1$ to the filter system at $z=0$. Note that uncertainties due to K+E corrections are not included in the error contours, and they purely show the uncertainty in the chi-squared best fitting.}
\label{fig:contours}
\end{figure}

\setcounter{table}{4}
\begin{table*}
\caption{Schechter parameters and luminosity densities for optical and NIR surveys in the literature}
\begin{tabular}{lccccccp{1.6cm}p{1.5cm}}
\hline \hline Reference &Band&Sample Size & $\lambda$ ($\mu
m$)&$\phi^{*}$ ($\rm h^3 \text{} Mpc^{-3}$)&$M^{*}$ (mag)&$\alpha$&j ($\times$
$10^{8}$ $h$ $L_{ \astrosun }$ $Mpc^{-3}$)&$\nu f_{\nu}$ ($10^{34}$
$h$ $W$ $Mpc^{-3}$)\\ \hline
\cite{tex:baldrysdssu} & $\rm u^{0.1}$& 43223&0.3224$\dagger$&0.0086 &-18.07$\ddag$&-1.05&2.28&1.80 \\
\cite{tex:blanton} & $\rm u^{0.1}$&113988&0.3224$\dagger$ & 0.0305 &-17.93&-0.92 &2.24& 1.77 \\
\cite{tex:sdssdr6} & $\rm u^{0.1}$&159018&0.3224$\dagger$& 0.0495 &-17.72&-1.05 &3.34 &2.64\\
\cite{tex:blanton} & $\rm g^{0.1}$&113988&0.4245$\dagger$ &0.0218 &-19.39&-0.89&1.75&3.63 \\
\cite{tex:sdssdr6} & $\rm g^{0.1}$ &256952&0.4245$\dagger$&0.0125 &-19.53&-1.10&1.29&2.67 \\
\cite{tex:blanton} & $\rm r^{0.1}$&113988& 0.5596$\dagger$&0.0149 &-20.44&-1.05 &1.85 &5.39\\
\cite{tex:sdssdr6} & $\rm r^{0.1}$ &466280&0.5596$\dagger$&0.0093 &-20.71&-1.26&1.78& 5.19\\
\cite{tex:blanton} & $\rm i^{0.1}$&113988& 0.6792$\dagger$ &0.0147 &-20.82&-1.00 & 2.11& 6.03\\
\cite{tex:sdssdr6} & $\rm i^{0.1}$ &461928&0.6792$\dagger$ &0.0114 &-20.93&-1.14&1.99 & 5.71 \\
\cite{tex:blanton} & $\rm z^{0.1}$&113988&0.8107$\dagger$ &0.0135 &-21.18&-1.08&2.71& 6.81 \\
\cite{tex:sdssdr6} & $\rm z^{0.1}$ &422643&0.8107$\dagger$ &0.0092 &-21.40&-1.26 &2.64&6.63 \\
\cite{tex:hj} & J&93841&1.250& 0.0071 &-22.85&-1.10 &2.97& 4.62 \\
\cite{tex:eke} & J&43553&1.250& 0.0139 &-22.39&-0.82 &3.29&5.11\\
\cite{tex:cole} & J&17173&1.250&0.0104 &-22.36&-0.96 &2.53&3.94 \\
\cite{tex:hj} & H &90317&1.644&0.0072 &-23.54&-1.11 &4.34&4.52 \\
\cite{tex:hj} & K &113988 &2.198&0.0074 &-23.83&-1.16 &5.86 &2.93\\
\cite{tex:cole} & K&17173 & 2.198&0.0108 &-23.44&-0.96 &5.20&2.60 \\
\cite{tex:kochanek} & K&4192& 2.198&0.0116 &-23.39&-1.09&5.46& 2.89 \\
\cite{tex:bell} & K&6282 & 2.198& 0.0143 &-23.29&-0.77 &5.58& 2.79 \\
\cite{tex:ajsmith} & K&36663& 2.198 &0.0176 &-23.17&-0.81&6.22&3.11 \\
\hline
\end{tabular}

$\dagger$ adjusted to effective filter rest wavelength for object at $z = 0.1$ (and then propagated through to the calculation of j and $\nu f_{\nu}$).\\
$\ddag$ adjusted to $h=1$.

\label{tab:othersurveys}
\end{table*}

\subsection{The Cosmic Spectral Energy Distributions}
We can calculate the luminosity density ($j_{\lambda}$) in each filter from the integration of the Schechter luminosity function, using the parameters listed in Table \ref{tab:magcutchanges} and Equation \ref{eqn:totlum}. We convert these measurements (in units of $hL_{\astrosun,\lambda}$Mpc$^{-3}$) to energy per frequency interval ($hW$Mpc$^{-3}\delta Hz$) which is more useful when comparing data with significantly differing filter widths and for comparison to model SEDs. To make this conversion we use:
\begin{equation}
\nu f(\nu)=\frac{c}{\lambda}\text{ }j_{\lambda}\text{ }10^{-0.4(M_{\astrosun,\lambda}-34.10)}
\end{equation}
where $\lambda$ is the effective wavelength, $j_{\lambda}$ is the total luminosity density and $M_{\astrosun,\lambda}$ is the absolute magnitude of the Sun in the specified filter. The constant value (34.10) derives from the definition of the AB magnitude scale (where 3631Jy equates to 0 mag; \citealt{tex:ab}). Note that the filter width technically should appear twice but cancels, i.e. to derive the total flux through the filter one should multiply by the filter width (in Hz), however to make a useful comparison it is more logical to show energy per $\delta$Hz which requires dividing by the filter width (in Hz). Our values for $j$ and $\nu f(\nu)$ are in the final two columns of Table \ref{tab:sdssmagcutchanges}, and values for previous surveys are in the final columns of Table \ref{tab:othersurveys}. \\ 
Figure \ref{fig:cosmice} shows the position of our total luminosity density values compared with previous measurements. The step-function seen between previous optical and NIR surveys is no longer apparent in our data (except for a slight decrement in $J$) perhaps suggesting that cosmic variance may indeed have played a part in this discrepancy. In general our data points are consistent with what has been published before but provides a relatively smooth distribution within a single survey suggesting that constructing the CSED from within a single survey volume is critically important.\\
\setcounter{figure}{6}
\begin{figure*}
\includegraphics[width=420pt]{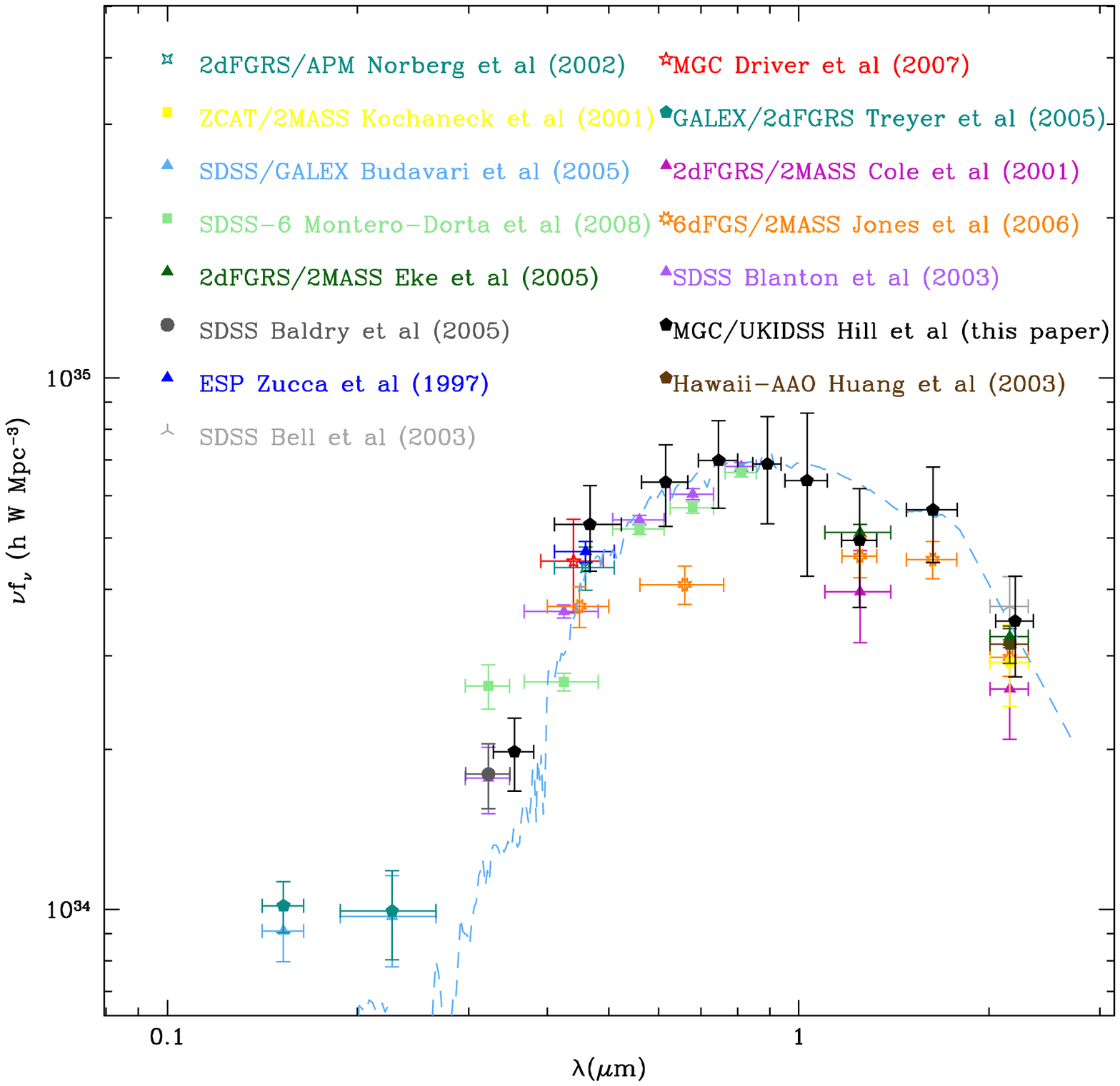}
\caption{Cosmic energy output from 0.1 to 3 $\mu m$. The model line shows the SED of a 13.2Gyr Sa-type galaxy from the
spectral template library of \citet{tex:pogg}. We also include datapoints from \citet{tex:baldrysdssu}, \citet{tex:bell}, \citet{tex:blanton}, \citet{tex:budavari}, \citet{tex:cole}, \citet{tex:dustatt}, \citet{tex:eke}, \citet{tex:hj},  \citet{tex:huang}, \citet{tex:kochanek}, \citet{tex:sdssdr6}, \citet{tex:pederapm}, \citet{tex:treyer} and \citet{tex:zucca}.}
\label{fig:cosmice}
\end{figure*}

\begin{figure}
\includegraphics[width=210pt]{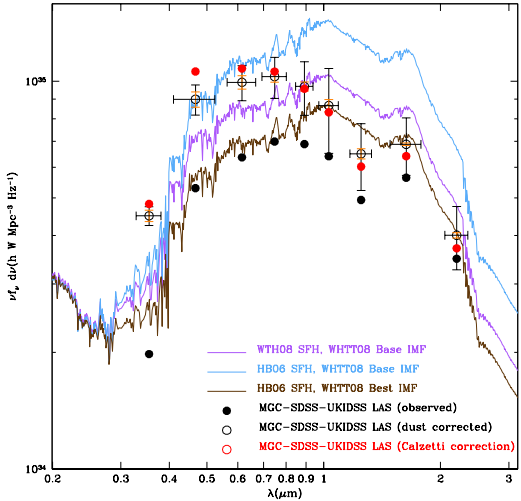}
\caption{The cosmic spectral energy distribution as derived from the  combined MGC-SDSS-UKIDSS LAS surveys compared to that expected from various cosmic star-formation history/initial mass function combinations as indicated (Wilkins priv. comm.). Solid symbols are the raw empirical data uncorrected for dust attenuation and open symbols are corrected for dust attenuation following \citealt{tex:uvopnir} (black), or \citealt{tex:calzettired} (red). Errors are only shown for the \citeauthor{tex:uvopnir} data for clarity, and are split into two components; the black errorbars show the uncertainty due to chi-squared fitting, the orange errorbars show the uncertainty due to k and e corrections.}
\label{fig:cosmicdc}
\end{figure}

The CSED values are still dust uncorrected and in Fig.~\ref{fig:cosmicdc} we show the pre- (solid symbols) and post- (open symbols) attenuated values. We adopt the prescription laid out in \citet{tex:uvopnir} to correct our cosmic SED; this results in corrections of  $\times2.27$, $\times1.69$, $\times1.56$, $\times1.47$, $\times1.41$, $\times1.35$, $\times1.32$, $\times1.22$, and $\times1.15$ in $u,g,r,i,z,Y,J,H,$ and $K$ respectively (black, solid symbols). For comparison, we also show dust corrections calculated using the prescription laid out in Section 3.3 of \citet{tex:calzettired}; using $E_{S}(B-V)=0.16$ (from the same paper), this results in corrections of $\times2.46$, $\times2.02$, $\times1.69$, $\times1.51$, $\times1.39$, $\times1.3$, $\times1.22$, $\times1.13$, and $\times1.06$ in $u,g,r,i,z,Y,J,H,$ and $K$ respectively (red, solid symbols). Overlaid are three expectations derived for us by Stephen Wilkins (priv. comm) from various cosmic SFH+IMF combinations using the PEGASE code. The blue curve is based on the cosmic star-formation history (CSFH) assembled by \citet{tex:beacom} from direct measurements reported in the literature but with a Salpeter IMF flattened below $0.5M_{\odot}$. The brown curve uses the same CSFH but adopts the best fitting IMF of \citet{tex:wilkins2}, which has a high-mass slope slightly shallower than the typical Salpeter value (i.e., $-2.15$ rather than $-2.35$).  \\  
Both appear to be inconsistent with our data, with the data fitting the blue curve in the optical and then tending towards the brown curve in the NIR. An intermediate solution could perhaps be found which would fit. The purple curve also adopts a Salpeter IMF but with the cosmic star-formation history derived by \citet{tex:wilkins}. This is based on constraints from the evolution of the total stellar mass history and which generally predicts a lower star-formation rate at higher redshift than reported in \citet{tex:beacom}. However, like the brown curve, the magenta curve fails to fit the data at shorter wavelengths. As we are in the process of assembling a much larger dataset with significantly smaller uncertainties (particularly diminishing the effects of cosmic variance) we defer a detailed comparison of the CSED, CSFR and IMF.\\
Here we conclude that our values are broadly consistent with the range of results reported in \citet{tex:wilkins2} and \citet{tex:wilkins}, and that refined measurements of the CSED should provide useful additional constraints on the CSFH and IMF. Such improvements should arise via the following measures:
\begin{enumerate}
\item A larger statistical sample
\item Deeper NIR photometry
\item Matched photometry/deblended solutions across all filters
\item A full bi-variate brightness distribution to model the selection bias
\item More sophisticated modelling of the $K(z)$ and $E(z)$ corrections
\end{enumerate}
These improvements are currently in progress within the ongoing Galaxy And Mass Assembly Survey (GAMA).\\

\section{Summary} \label{section:conclude}
We have combined data from the MGC, SDSS and UKIDSS LAS to produce a master catalogue of $\sim 10$k objects with $ugrizYJHK$ photometry and 100 per cent redshift completeness. After careful consideration of the colour bias and restricting our sample to $z<0.1$ we produce nine pseudo-flux limited samples comprising of between 3328 to 1589 galaxies, depending on the filter. Using a modified SWML method and least-$\chi^2$ fitting to a Schechter function parametrisation, we recover luminosity distributions and functions in all nine filters. The resulting LFs in the optical typically produce luminosity functions consistent with previous publications from the SDSS collaboration albeit with slightly higher normalisations. In the NIR we recover significantly distinct Schechter funtion parameters which we attribute to the rapid improvement in NIR technologies and fields-of-view. Finally we derive the cosmic spectral energy distribution from $u$ to $K$ which is shown to be smooth. After correcting for dust attenuation we find that our observed CSED agrees within the errors with that expected and in due course these observations should provide useful new constraints on both the cosmic star-formation history and the universality of the initial mass function. Work is currently underway to significantly improve upon the results reported here via the Galaxy And Mass Assembly (GAMA) survey.

\section{Acknowledgements} 
We would like to acknowledge the help Mike Read gave in modifying the WFCAM interface, making the extraction of a large number of UKIDSS images a much easier task. We would also like to acknowledge Paul Hewett, for providing us with accurate values of $M_{\astrosun,\lambda}$ for both the UKIDSS and SDSS filter sets, and Antonio Montero-Dorta for providing us with his number count dataset so that we could compare our sample to his. DTH is funded by a STFC studentship. EC is funded by a STFC grant. SPD acknowledges support from the Australian National Telescope Facility and the University of Western Australia as a visiting Professor.\\ 

The Millennium Galaxy Catalogue consists of imaging data from the Isaac Newton Telescope and spectroscopic data from the Anglo Australian Telescope, the ANU 2.3m, the ESO New Technology Telescope, the Telescopio Nazionale Galileo and the Gemini North Telescope. The survey has been supported through grants from the Particle Physics and Astronomy Research Council (UK) and the Australian Research Council (AUS). The data and data products are publicly available from http://www.eso.org/$\sim$jliske/mgc/ or on request from J. Liske or S.P. Driver.\\ 

Funding for the SDSS and SDSS-II has been provided by the Alfred P. Sloan Foundation, the Participating Institutions, the National Science Foundation, the U.S. Department of Energy, the National Aeronautics and Space Administration, the Japanese Monbukagakusho, the Max Planck Society, and the Higher Education Funding Council for England. The SDSS Web Site is http://www.sdss.org/. The SDSS is managed by the Astrophysical Research Consortium for the Participating Institutions. The Participating Institutions are the American Museum of Natural History, Astrophysical Institute Potsdam, University of Basel, University of Cambridge, Case Western Reserve University, University of Chicago, Drexel University, Fermilab, the Institute for Advanced Study, the Japan Participation Group, Johns Hopkins University, the Joint Institute for Nuclear Astrophysics, the Kavli Institute for Particle Astrophysics and Cosmology, the Korean Scientist Group, the Chinese Academy of Sciences (LAMOST), Los Alamos National Laboratory, the Max-Planck-Institute for Astronomy (MPIA), the Max-Planck-Institute for Astrophysics (MPA), New Mexico State University, Ohio State University, University of Pittsburgh, University of Portsmouth, Princeton University, the United States Naval Observatory, and the University of Washington.\\

The UKIDSS project is defined in \citealt{tex:ukirt}. UKIDSS uses the UKIRT Wide Field Camera (WFCAM; \citealt{tex:casali}) and a photometric system described in \citealt{tex:hewett}. The pipeline processing and science archive are described in Irwin et al (2008) and \citealt{tex:wfcam}.\\

STILTS is a set of command-line based catalogue matching tools, based upon the STIL library. It is currently being supported by its author (Mark Taylor), and can be downloaded from http://www.star.bris.ac.uk/$\sim$mbt/stilts/.

\bibliography{mgc20_spd_v7}

\appendix \section{Problems with CASU source extraction software} \label{subsec:deblend} The WFCAM Science Archive (WSA,
\citealt{tex:wfcam}) is the storage facility for post-pipeline, calibrated UKIDSS data. It provides users with access to fits images and CASU-generated object catalogues for all five UKIDSS surveys, along with copies of catalogues from a number of other surveys, including the SDSS and the MGC. The creators of the archive have also included cross matched data tables that link objects in the different catalogues. The LAS-MGC cross match table ($lasSourceXmgcDetection$) contains the ID numbers for all MGC objects within 10 arcsec of a LAS source, and the ID of that LAS source. Using this tool, we downloaded catalogues matching MGC-Bright $B$ band luminosities to their counterpart LAS $Y$, $J$, $H$ and $K$ band luminosities. We required objects where both the MGC and LAS objects were definitely galaxies (MGC $CLASS=1$ and LAS $pGalaxy > 0.9$; this criteria is only used here to guarantee a galaxy-only sample, and is not used where completeness is important), not in an MGC exclusion region ($INEXR = 0$), with good photometry ($QUALITY \le 2$) and with no major errors in the LAS observation ($ppErrBits < 256$; following \citealt{tex:ajsmith}, this criteria removes all objects that lie within dither offsets, all possible crosstalk artifacts, objects with bad pixels and objects close to saturation). We also specified that the distance between the galaxy's centroid in the two surveys had to be no more than $2.5$ arcsec apart ($lasSourceXmgcDetection.distanceMins \le 2.5/60$). We define $K_{\rm UKIDSS}$ as the $K$ band Petrosian magnitude ($lasSource.kPetroMag$) taken from the WSA dataset.\\
\setcounter{figure}{0}
\begin{figure*}
\includegraphics[]{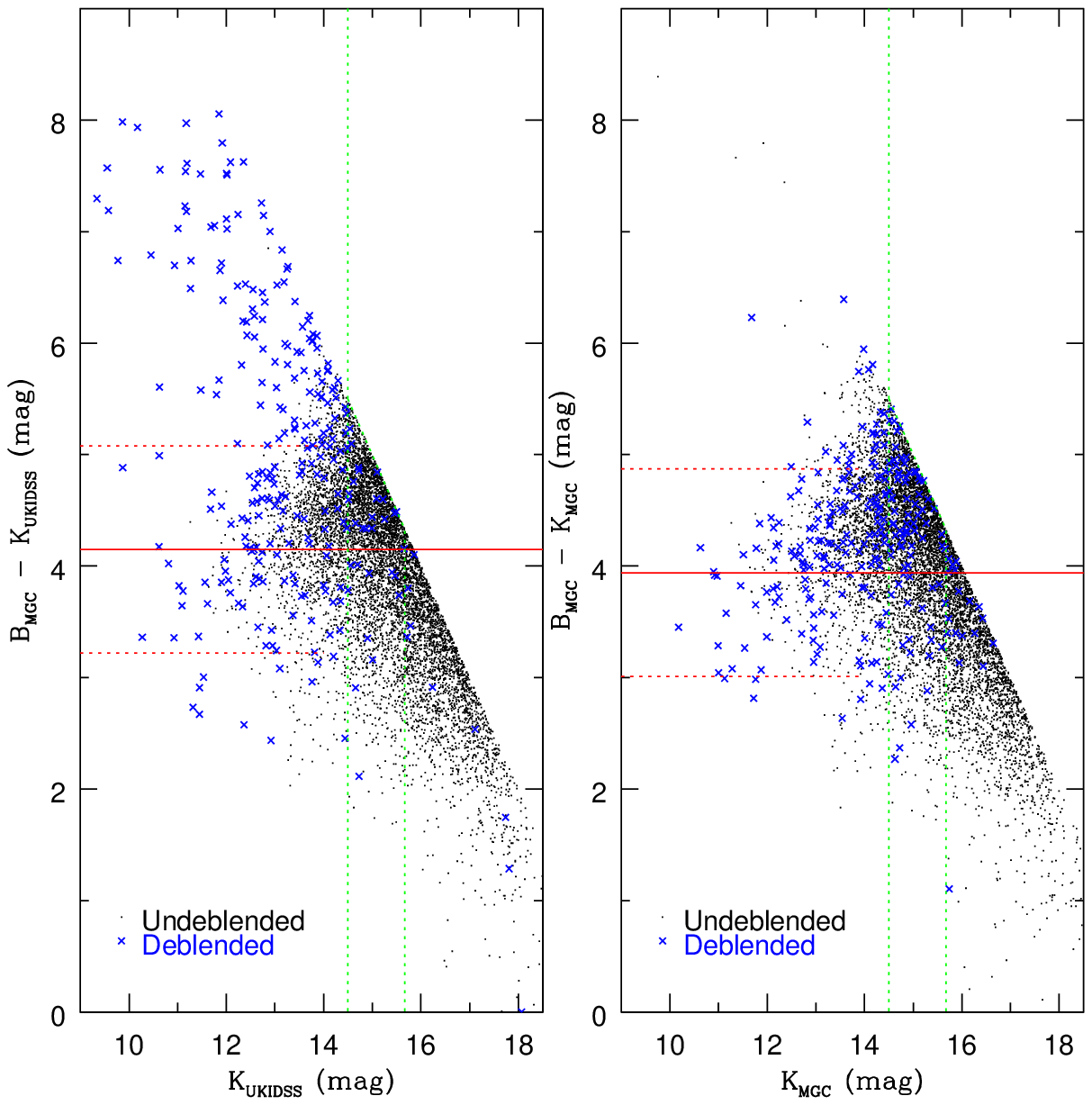}
\caption{Comparisons between the MGC B-K values for K values taken from the UKIDSS Survey data ($K_{\rm UKIDSS}$), and from Sextracted UKIDSS images ($K_{\rm MGC}$). The dotted green lines are the apparent magnitude cuts used for the $K$ band sample, and the red lines are the colour median and colour outlier lines for each sample.}
\label{fig:sexbk}
\end{figure*}
Unfortunately, a problem arises when the colour distribution of the sample is plotted (see the left image of Figure \ref{fig:sexbk}). There are a significant number of very bright objects that are extremely red. \cite{tex:ajsmith} noted this issue and have traced this problem to a catastrophic fault in the CASU deblending algorithm, which is causing deblended galaxies to become significantly brighter than their parent object, in some cases by several mag. This is highlighted in the left hand image of Figure \ref{fig:sexbk}; while the undeblended bright galaxies (black points) are clustered around the $B_{\rm MGC} - K_{\rm UKIDSS} = 4$ mag line, the deblended bright galaxies (blue crosses) are spread across a much wider colour range. \\ 
Unfortunately, the pre-deblended data are not output by the pipeline (when the CASU source extractor breaks up brighter galaxies, it does not generate any parameters for the parent object; Irwin et al, in preparation), so this fault is not trivial to correct or quantify. It is possible to remove the deblended galaxies from the sample by filtering upon an error bit designated as the deblender flag ($lasSource.kppErrBits \& 0\rm{x}00000010 = 0$) but this would leave us with a biased dataset. This option was adopted by \cite{tex:ajsmith}, but Figure \ref{fig:fdeb} shows our reservation; although only 3.5 per cent of galaxies in the unlimited $K$ band sample, 3.2 per cent in $H$, 3.1 per cent in $J$ and 10.8 per cent in $Y$ are deblended, they are not uniformly distributed across the apparent magnitude range.\\
\begin{figure}
\includegraphics[width=220pt, clip=true, trim=10 280 300 0 ]{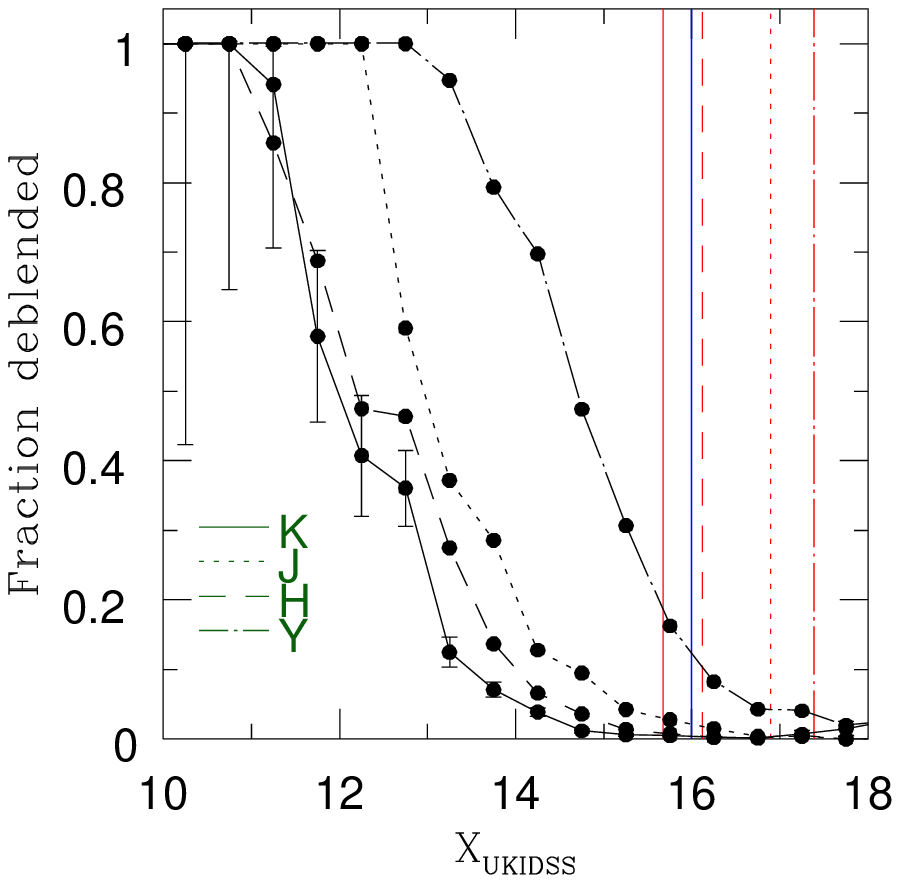}
\caption{The fraction of galaxies deblended as a function of $X_{\rm
  UKIDSS}$, where $X$ is $Y$, $J$, $H$ or $K$. The red vertical
lines are our band dependent apparent magnitude limits, the blue
vertical line is the $K$ band apparent magnitude limit of
\citet{tex:ajsmith}.}
\label{fig:fdeb}
\end{figure}
We initially attempted to overcome the problem by using the SDSS optical colours to predict the $Y$, $J$, $H$ and $K$ band fluxes for these problem galaxies. Unfortunately, the correlation was too noisy ($\Delta m \sim \pm 1$mag), so we were forced to abandon the CASU DR3PLUS catalogue products provided through the archive and entirely re-derive all $Y$, $J$, $H$ and $K$ photometry from the reduced fits images.

\subsection{Reanalysis of UKIDSS data using Sextractor} \label{subsec:seanalysis}
\begin{figure}
\includegraphics[width=220pt]{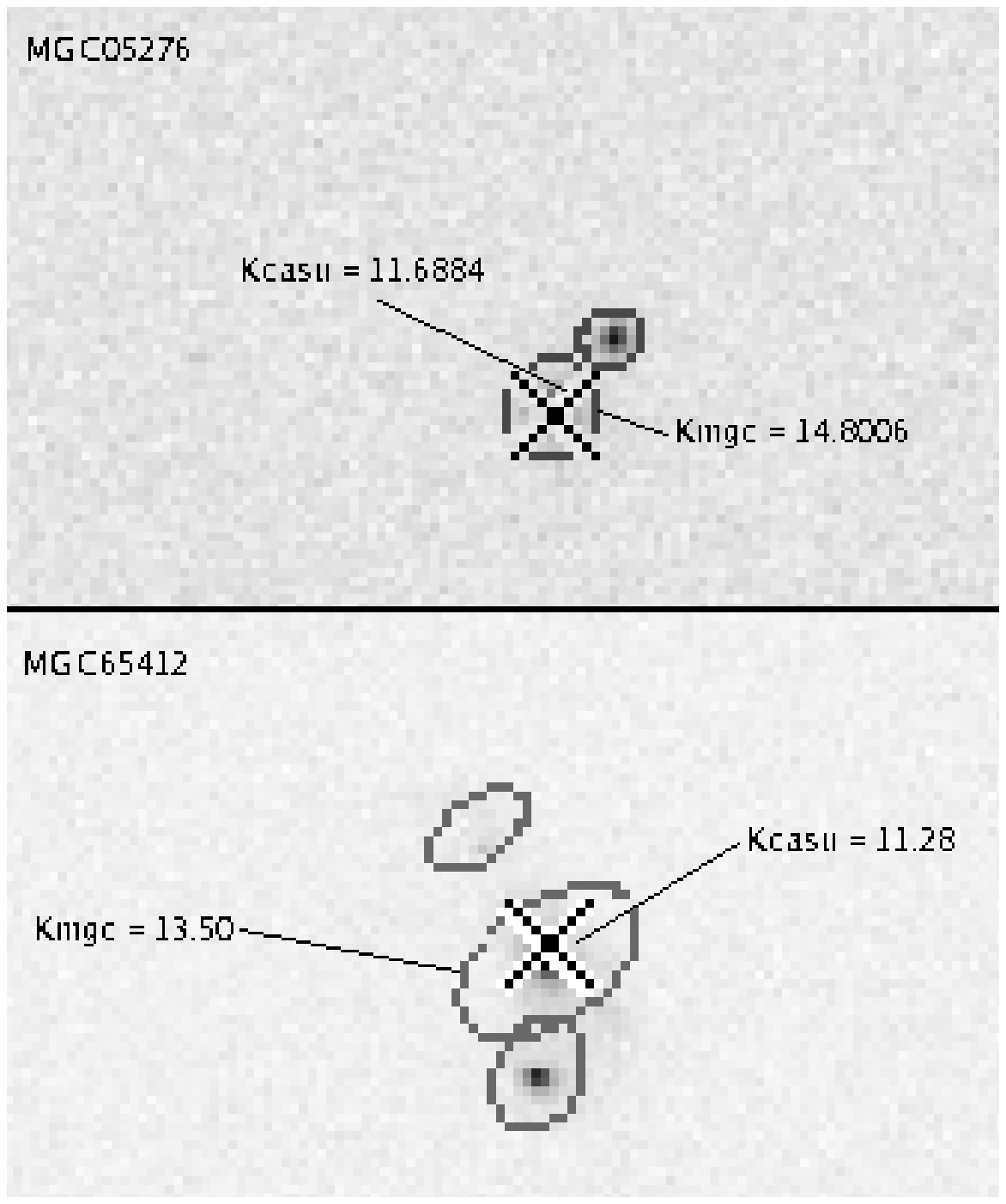}
\caption{The position of the $K_{\rm MGC}$ Kron aperture (grey ellipse) and the central position of the $K_{\rm UKIDSS}$ galaxy (black cross) for the deblended galaxies MGC65412 and MGC05276}
\label{fig:twodeb}
\end{figure}

The WSA contains a tool ($MultiGetImage$) for extracting $1$ arcminute $\times$ $1$ arcminute fits images of 500 objects when given a list of their coordinates. The RA and Dec of the MGC-Bright galaxies were input into this program and images were extracted (as was a list of the galaxies where no image was available because of limited coverage).\\ 
The SExtractor object extraction program \citep{tex:sext} was used to extract objects from these images (taking zeropoint, pixel size, gain, seeing and background levels from the UKIDSS fits file headers), and the fluxes and aperture centres for objects in each image catalogued. Both elliptical aperture Petrosian and \textsc{Best} fluxes were calculated for each object (the SExtractor \textsc{Best} flux generally uses the Kron flux, except in crowded regions where it uses a corrected isophotal flux). As $B_{\rm MGC}$ magnitudes were also calculated using \textsc{Best} apertures, we adopt \textsc{Best} magnitudes for our NIR magnitudes (we define $K_{\rm MGC}$ as the SExtracted-\textsc{Best} magnitude and $K_{\rm MGC,Petrosian}$ as the SExtracted-Petrosian magnitude).\\ 
Distances between the position of each newly SExtracted object and the centre of that image were calculated (using the catalogued aperture centres and axis sizes taken from fits file headers). As all extracted images were centred on the position of an MGC-Bright galaxy, the object that was extracted closest to the centre of each frame is assumed to be the match to that MGC galaxy. The apparent magnitude of this object is calculated from its flux ($F$), exposure time ($t$), extinction ($Ext$), zeropoint ($Z_{\rm pt}$) and airmass ($\rm{sec} \chi_{\rm mean}$; the former is taken from the SExtractor created catalogue, the rest are taken from UKIDSS fits file headers), using:
\begin{equation} \label{eqn:apmag}
m = Z_{\rm pt} - 2.5 \rm{log}_{\rm 10}( \frac{\textit{F}}{\textit{t}} )- \rm{Ext.}  (\rm{sec} \chi_{\rm mean} - 1) 
\end{equation}
Again, we exclude all NIR objects that are $> 2.5$ arcsec from the MGC galaxy's centre. The right hand graph in Figure \ref{fig:sexbk} show how the $B_{\rm MGC}$-$K_{\rm MGC}$ distribution compares with that from WSA dataset. The online archive contains 58 deblended galaxies with $B_{\rm MGC}-K_{\rm UKIDSS} > 6$ mag. When re-extracted, only 2 of those galaxies have $B_{\rm MGC} -K_{\rm MGC}$ in that range (one is a distant galaxy that will be removed from our luminosity function when we impose redshift limits and the other is just a very red galaxy). Figure \ref{fig:twodeb} contains K band images of two of the deblended MGC galaxies, the position of their $K_{\rm MGC}$ and $K_{\rm UKIDSS}$ apertures and the luminosity returned using each method. Figure \ref{fig:ukvsmgc} shows how $K_{\rm UKIDSS}$, $K_{\rm MGC}$ and $K_{\rm MGC,Petrosian}$ photometry compares. We note that the dominant error in galaxy photometry is typically due to the flux measurement method. We find for an individual galaxy a typical uncertainty of $\pm 0.04$mag between the elliptical aperture \textsc{BEST} and Petrosian methods, and therefore adopt this as representative of our standard luminosity error.

\setcounter{figure}{3}
\begin{figure*}
\includegraphics[]{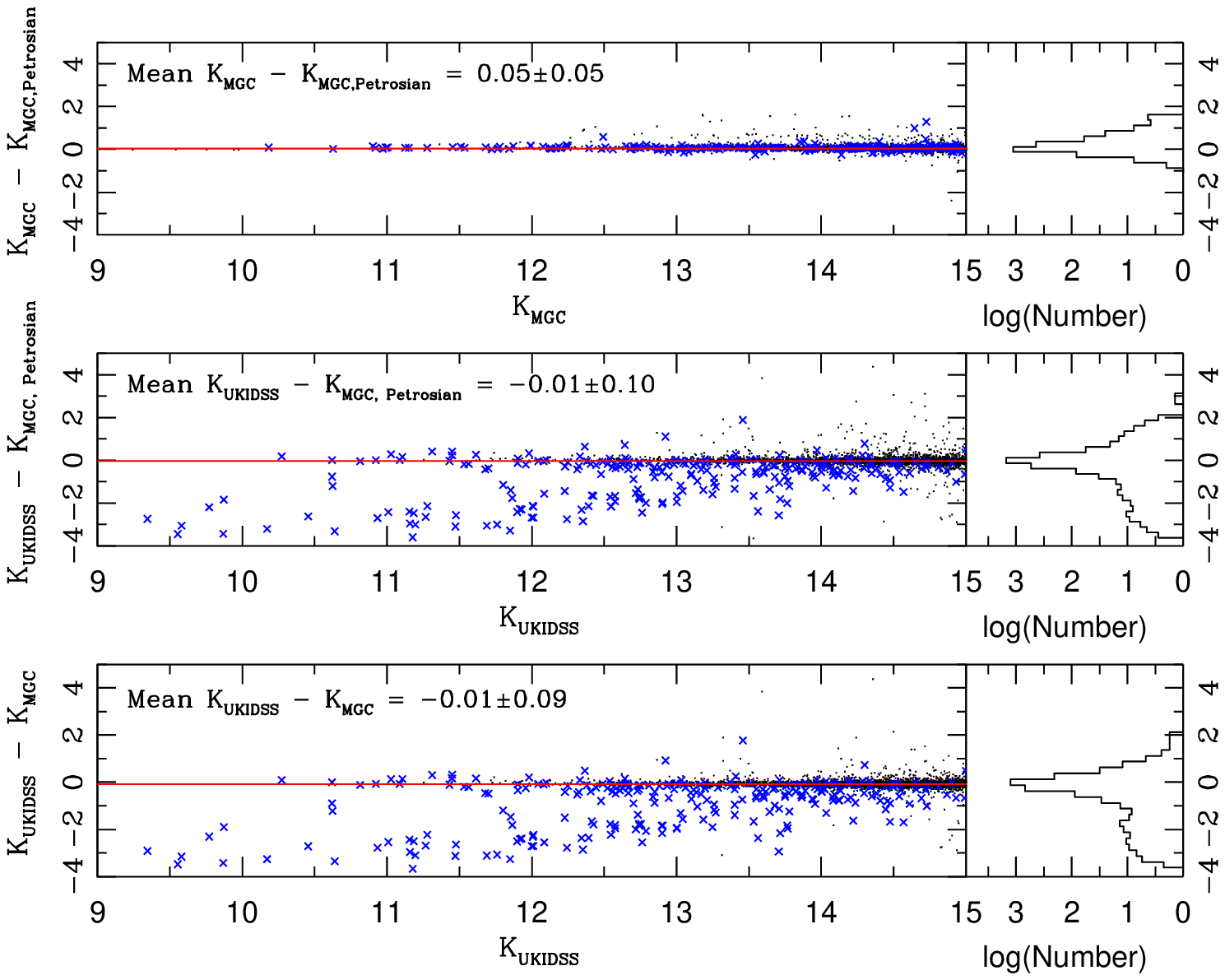}
\caption{Comparisons between the $K$ band galaxy luminosities using magnitudes taken from the UKIDSS Survey data ($K_{\rm UKIDSS}$), and from SExtracted UKIDSS images using circular Petrosian ($K_{\rm MGC,Petrosian}$) and \textsc{Best} ($K_{\rm MGC}$) methods. As in Figure \ref{fig:sexbk}, blue crosses are galaxies that have been flagged as deblended in the UKIDSS survey data, and black dots are those that have not.}
\label{fig:ukvsmgc}
\end{figure*}

\subsubsection{Near-IR galaxies with no MGC match} \label{subsec:nomatch}
It is possible that there are galaxies in the UKIDSS survey that have no match to the MGC, despite being inside its area of coverage. While some of these objects may be misclassified by the automated classification system, a number of these objects could potentially be particularly red galaxies (e.g. heavily dust attenuated or exceptionally high redshift galaxies) that have slipped below the $B_{\rm MGC}=20$ mag threshold and must be recovered for the calculation of the appropriate NIR luminosity function. We extracted fits files for all UKIDSS galaxies ($pGal > 0.9$) with $X < B_{\rm MGC, threshold} - (B_{\rm MGC, threshold}-X_{\rm Survey})_{\rm median}$. We used SExtractor to extract objects from these images (using the same parameters as in subsection \ref{subsec:seanalysis}), and calculated their luminosity using the \textsc{Best} aperture. We excluded all objects that were not covered by MGC CCDs. Following \cite{tex:mgc}, we also excluded all objects with $stellaricity > 0.98$. The remaining objects that lie above our apparent magnitude limits (see subsection \ref{subsec:apmlimit}) were checked by eye, to remove any that looked like they were noise (or were otherwise suspect) that had been misclassified. We found a total of 100 NIR galaxies without a MGC match brighter than our $K$ band apparent magnitude limit, 76 in $H$, 73 in $J$ and 42 galaxies brighter than our $Y$ band sample. However, as these galaxies are still below our $B$ magnitude limit, they are not used in the calculation of our luminosity function.

\setcounter{figure}{4}
\begin{figure*}
\includegraphics[width=440pt]{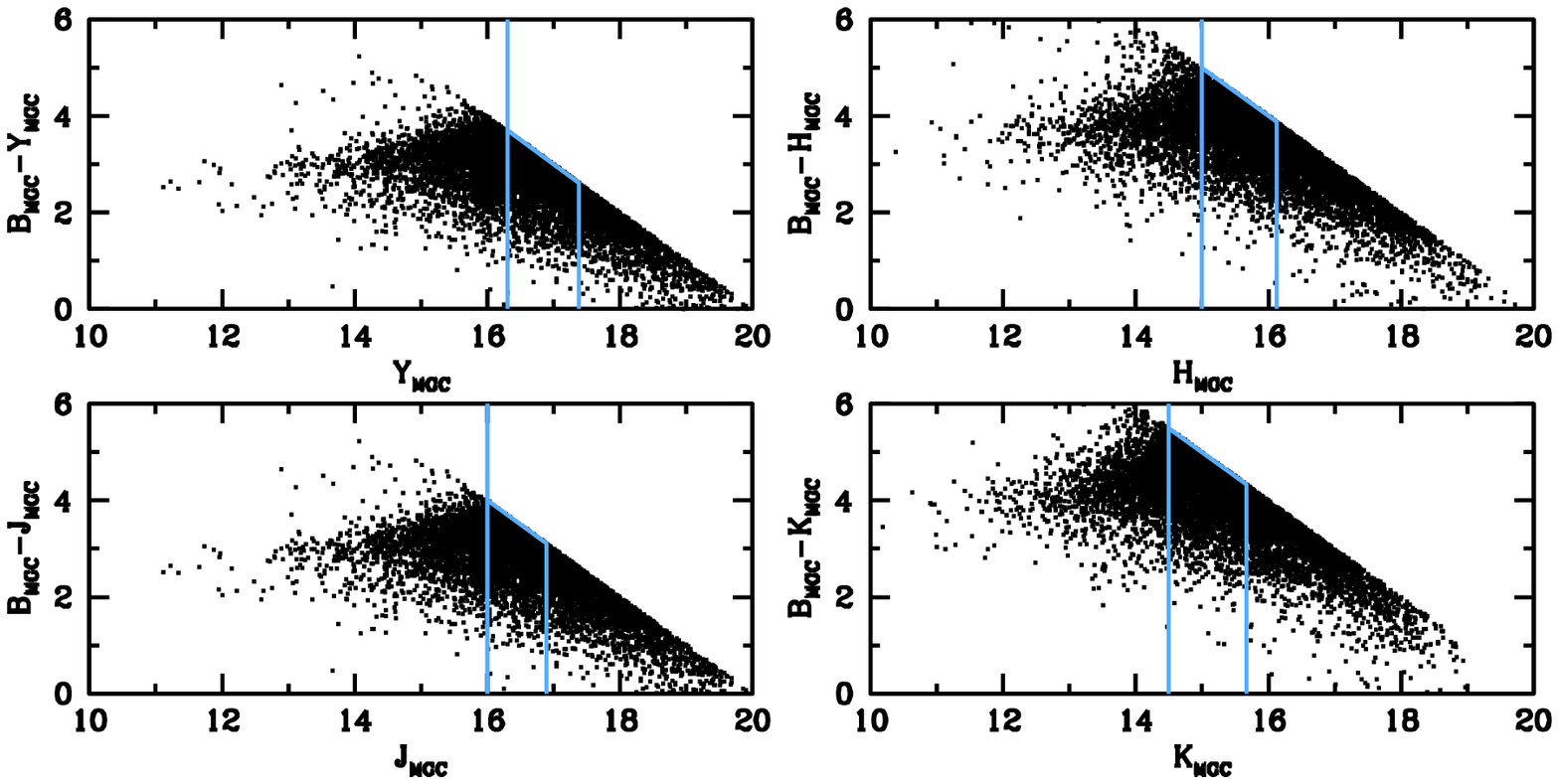}
\caption{$B_{\rm MGC}$,$B_{\rm MGC}-X_{\rm MGC}$ and $X_{\rm MGC}$,$B_{\rm MGC}-X_{\rm MGC}$ graphs for $X$=$Y$, $J$, $H$ and $K$ band data. The vertical lines in the left hand graphs are the apparent magnitude cuts. All galaxies above these lines are excluded from our luminosity function samples.}
\label{fig:nocut}
\end{figure*}

\subsubsection{Colour outliers} \label{subsec:colout}
Whilst our new catalogues produced by SExtractor contain a smaller number of colour outliers, some still remain and can be seen at the extremes in Figure \ref{fig:nocut}). The median and standard deviation of $B-X$ was calculated, and we define colour outliers to have a colour $> \rm{Median} + 3 \sigma$ or $< \rm{Median} - 3 \sigma$. We have manually reexamined these objects and, in 11 galaxies in the $K$ band, 19 in $H$, 1 in $J$ and 1 in $Y$, the program is incorrectly deblending an object. In these cases we modify the SExtractor deblending parameters to achieve a consistent deblending outcome.

\label{lastpage}
\end{document}